\documentclass[preprint]{emulateapj}

\slugcomment{Draft July 2008}

\shorttitle{NGC 2770 - a SN Ib factory?}
\shortauthors{C. C. Th\"one et al.}

\begin{document}

\title{NGC 2770 - a supernova Ib factory? \footnote{Based on observations with the Nordic Optical Telescope, ESO proposal 080.D-0526, the GALEX and NED databases}}

\author{Christina C. Th\"one \altaffilmark{1}, Micha\l{} J. Micha\l owski \altaffilmark{1}, Giorgos Leloudas \altaffilmark{1}, Nick L. J. Cox \altaffilmark{2}, Johan P. U. Fynbo \altaffilmark{1}, Jesper Sollerman \altaffilmark{1,3}, Jens Hjorth \altaffilmark{1} and Paul M. Vreeswijk \altaffilmark{1}}
\altaffiltext{1}{Dark Cosmology Centre, Niels-Bohr-Institute, University of Copenhagen, Juliane Maries Vej 30, 2100 K\o benhavn \O, Denmark}
\altaffiltext{2}{Herschel Science Centre, European Space Astronomy Centre, ESA, P.O.Box
78, E-28691 Villanueva de la Ca\~nada, Madrid, Spain}
\altaffiltext{3}{Stockholm Observatory, Alba Nova, 10691 Stockholm, Sweden}
\email{cthoene@dark-cosmology.dk}

\begin{abstract}
NGC 2770 has been the host of three supernovae of Type Ib during the last 10 years, SN 1999eh, SN 2007uy and SN 2008D. SN 2008D attracted special attention due to the serendipitous discovery of an associated X-ray transient. In this paper, we study the properties of NGC 2770 and specifically the three SN sites to investigate whether this galaxy is in any way peculiar to cause a high frequency of SNe Ib. We model the global SED of the galaxy from broadband data and derive a star-formation and SN rate comparable to the values of the Milky Way. We further study the galaxy using longslit spectroscopy covering the major axis and the three SN sites. From the spectroscopic study we find subsolar metallicities for the SN sites, a high extinction and a moderate star-formation rate. In a high resolution spectrum, we also detect diffuse interstellar bands in the line-of-sight towards SN 2008. A comparison of NGC 2770 to the global properties of a galaxy sample with high SN occurance ($\geq$ 3 SN in the last 100 years) suggests that NGC 2770 is not particularly destined to produce such an enhancement of observed SNe observed. Its properties are also very different from gamma-ray burst host galaxies. Statistical considerations on SN Ib detection rates give a probability of $\sim$ 1.5\% to find a galaxy with three Ib SNe detected in 10 years. The high number of rare Ib SNe in this galaxy is therefore likely to be a coincidence rather than special properties of the galaxy itself. NGC 2770 has a small irregular companion, NGC 2770B, which is highly starforming, has a very low mass and one of the lowest metallicities detected in the nearby universe as derived from longslit spectroscopy. In the most metal poor part, we even detect Wolf-Rayet features, against the current models of WR stars which require high metallicities.
\end{abstract}


\section{Introduction}

Massive stars end their lives in various ways, as governed by their mass, composition, angular momentum and whether or not they are interacting with a companion star \citep[e.g.][]{Heger02}. The most massive stars loose their outer layers through winds whose strength strongly depend on the metallicity of the star \citep[e.g.][and references therein]{Crowther02}. These stripped stars explode as supernovae (SNe) Type Ib (SN Ib) if they have lost their hydrogen envelope or as SNe Type Ic (SN Ic) if they have also lost their He envelope. If the star is also rapidly rotating, it is believed that the star might even produce a Gamma-Ray Burst (GRB) \citep{MacFadyen99, Woosley06}. SNe Ib/c are much rarer than SNe II, which are produced by stars with masses of 8 to 40 M$_\odot$, whereas SNe Ib/cs are assumed to require zero age main sequence (ZAMS) masses of $\gtrsim$ 35 M$_\odot$ for non-rotating stars \citep{Woosley02}. GRBs are even less common than SNe Ic which is in line with the general picture that GRB progenitors require some special conditions to produce a GRB. 

Only for SNe II has it been possible to identify the progenitor star  - which turned out to be yellow, red or blue giants with masses of around 8 -- 20 M$_\odot$, e.g. SN 1987A \citep{Gilmozzi87}, SN 2002ov \citep{Li07}, SN 2004A \citep{Hendry06} and SN 2006gl \citep{Gal-Yam07}. Furthermore, several detections have been made using preexplosion imaging from the {\it HST} archive \citep[e.g.][]{Smartt03}. SNe Ib/c have so far evaded an identification with a progenitor star \citep{Maund05, Crockett07, Crockett08}. This is also the case for GRBs which most often occur at distances that do not even allow us to resolve their host galaxies.

When the progenitor cannot be directly identified, studying the environment can provide important information about what kind of star exploded.  Spatially resolved photometric studies of the sites of
different types of SNe show that SNe Type II trace the light of their host galaxies while GRBs are more concentrated towards bright regions \citep{Fruchter06} and that SNe Type Ib, Ic and GRBs appear to be differently distributed within their hosts \citep{Kelly07}. 
Spectroscopic investigations show that the broadlined SNe Ic which are not connected to GRBs are found at sites with higher metallicities than those that are connected to GRBs \citep{Modjaz08a}. A spatially resolved study of the host of one
long-duration GRB which was not connected to a SN also showed a low metallicity in comparison to the rest of the host galaxy
\citep{Thoene08}.

Long duration GRBs have been found to be associated with broadlined SNe Ic. There are four spectroscopically confirmed cases so far, namely GRB\,980425 \citep{Galama98}, GRB\,030329 \citep{Hjorth03, Matheson03, Stanek03}, GRB\,031203 \citep{Cobb04, Thomson04, Malesani04, Gal-Yam04} and GRB\,060218 \citep{Pian06,Sollerman06, Modjaz06, Mirabal06}. For all other nearby long GRBs up to 2006 where a SN could have been observed, additional light from the SN component was found in the late time lightcurves \cite[e.g.,][]{Zeh04}. GRBs therefore offer a unique opportunity to observe a SN from the very onset of the explosion, as indicated by the prompt emission from the GRB. The connection between GRBs and broadlined SNe Ic, however, had to be revised in 2006 when two long-duration GRBs were found not to show any sign of an associated 
SN \citep{Fynbo06, Gehrels06, DellaValle06, Gal-Yam06, Ofek07}.

One important supernova that recently caught the attention of the community is 
SN 2008D which was associated with an X-ray transient (XT) \citep{Soderberg08}. This supernova was serendipitously detected while the XRT instrument onboard the {\emph Swift} satellite \citep{Gehrels04} observed another supernova, SN 2007uy, in the same galaxy. Whether this prompt XT originated from a weak GRB-like event or if it was due to the shock breakout from the star is still under debate \citep{Soderberg08, Modjaz08b, Malesani08, Xu08}. The X-ray emission, however, is clearly associated with the
onset of the explosion, and consequently SN 2008D was one of the earliest observed SNe. The SN spectrum evolved from a smooth spectrum with small undulations characteristic of a high-velocity ejecta into a typical SN Ib \citep{Soderberg08,
  Modjaz08b, Malesani08}. Also the host galaxy of this supernova, NGC 2770, has attracted some attention as it produced three 
SNe within the last 9 years, SNe 1999eh, 2007uy and 2008D. Intriguingly, all three were stripped envelope core-collapse SNe Ib.

In this paper, we present a study on the local properties at the SN sites as well as of other regions in NGC 2770 with longslit spectroscopy and compare the host itself with a sample of other supernova producing galaxies. We want to investigate if the occurrence of three recent SNe Ib can be explained by some physical properties of the host galaxy. In Section 2 we present the
observations and data reduction of the spectra, Section 3 studies the global properties of NGC 2770 and properties derived from modeling the spectral energy distribution (SED). In Section 4 we analyze the different regions in the host in terms of metallicity, extinction and star formation rate (SFR). Section 5 compares NGC 2770 to a sample of other nearby galaxies with several recent SNe and discuss the probability to find three SNe Ib in a galaxy within nine years.  Finally, Section 6 investigates the properties of the companion galaxy NGC 2770B. Throughout the paper we use a cosmology with H$_0$=70 km s$^{-1}$ Mpc$^{-1}$, $\Omega_\Lambda$=0.7 and $\Omega_\mathrm{m}$=0.3. At z = 0.007 this corresponds to 0.13 kpc per arcsecond and we use a distance of 27 Mpc to NGC 2700.

\section{Observations}\label{NGC2770:observations}

We carried out optical spectroscopy using the FORS2 spectrograph at the VLT on Jan. 11.31, 2008. A 1\farcs0 arcsec wide slit together with the 300V grism provided a resolution of 11 \AA{}. The slit covered the site of SN 2008D as well as some outer parts of the galaxy. We obtained further spectra with ALFOSC at the NOT on Jan. 13, 15 and on Feb. 03 using grism 4. These observations were also part of the observations reported in \cite{Malesani08}. While \cite{Malesani08} used these data to study the supernova itself, we instead concentrate on the host galaxy. The three NOT spectra were obtained at three different slit positions (see also Fig. \ref{NGC2770:slitmetallicities}). The first included the positions of SNe 2008D and 2007uy which were also detected in the spectrum. The second one was placed along the major axis of the galaxy and the last one included the positions of SNe 2008D and 1999eh, where the latter one had already faded. On Jan. 15.21, we also obtained one 600s spectrum of the nearby galaxy NGC 2770B with ALFOSC \citep{FynboNGC2770GCN}.  Reduction and flux calibration of the spectra were done using standard tasks in IRAF. For the flux calibration we used the standard stars HD 19445, BD+75325 and BD+332642 for the ALFOSC spectra and  GD 108 for the FORS2 spectrum. From the longslit spectra we then extracted traces of equal sizes at the positions of the SNe and other HII regions in the galaxy (see Section \ref{NGC2770:parts})

Furthermore, we obtained high-resolution echelle spectra of SN 2008D with UVES \citep{Dekker00} at the VLT on Jan. 18, 2008. We used the DIC1 (390+564) setting together with a 1\farcs0 wide slit. This set-up covers a wavelength  range from 3300 to 6650\AA{} at a resolution of 6--8 km~s$^{-1}$.  The UVES data have been reduced and extracted using the ESO CPL pipeline \footnote{http://www.eso.org/sci/data-processing/software/pipelines/  uves/uves-pipe-recipes.html} and flux-calibrated with the master
response curves provided by ESO \footnote{http://www.eso.org/observing/dfo/quality/UVES/qc/re- sponse.html}. 

We also obtained imaging data with FORS1 at the VLT on March 16 using an H$\alpha$ filter shifted to z$=$ 0.007 with a central wavelength of 6604 \AA{} and a width of 64 \AA{} as well as an offband filter for which we used the H$\alpha$ filter at z$=$0 with $\lambda_\mathrm{center} =$ 6563 \AA{} and $\Delta \lambda =$ 61 \AA{}. In order to flux calibrate the H$\alpha$ image we determined the flux within the H$\alpha$ line from one of the traces extracted from the FORS spectra. We compared this flux with the counts within several rectangular apertures of the same size as for the extraction of the spectrum (1 $\times$ 3 arcsec) at the same positions along the slit by accounting for the different continuum levels from the galaxy and the varying strength of the [N\,{\sc ii}] line which is included in the H$\alpha$ narrowband filter.

\begin{figure}
\includegraphics[width=\columnwidth]{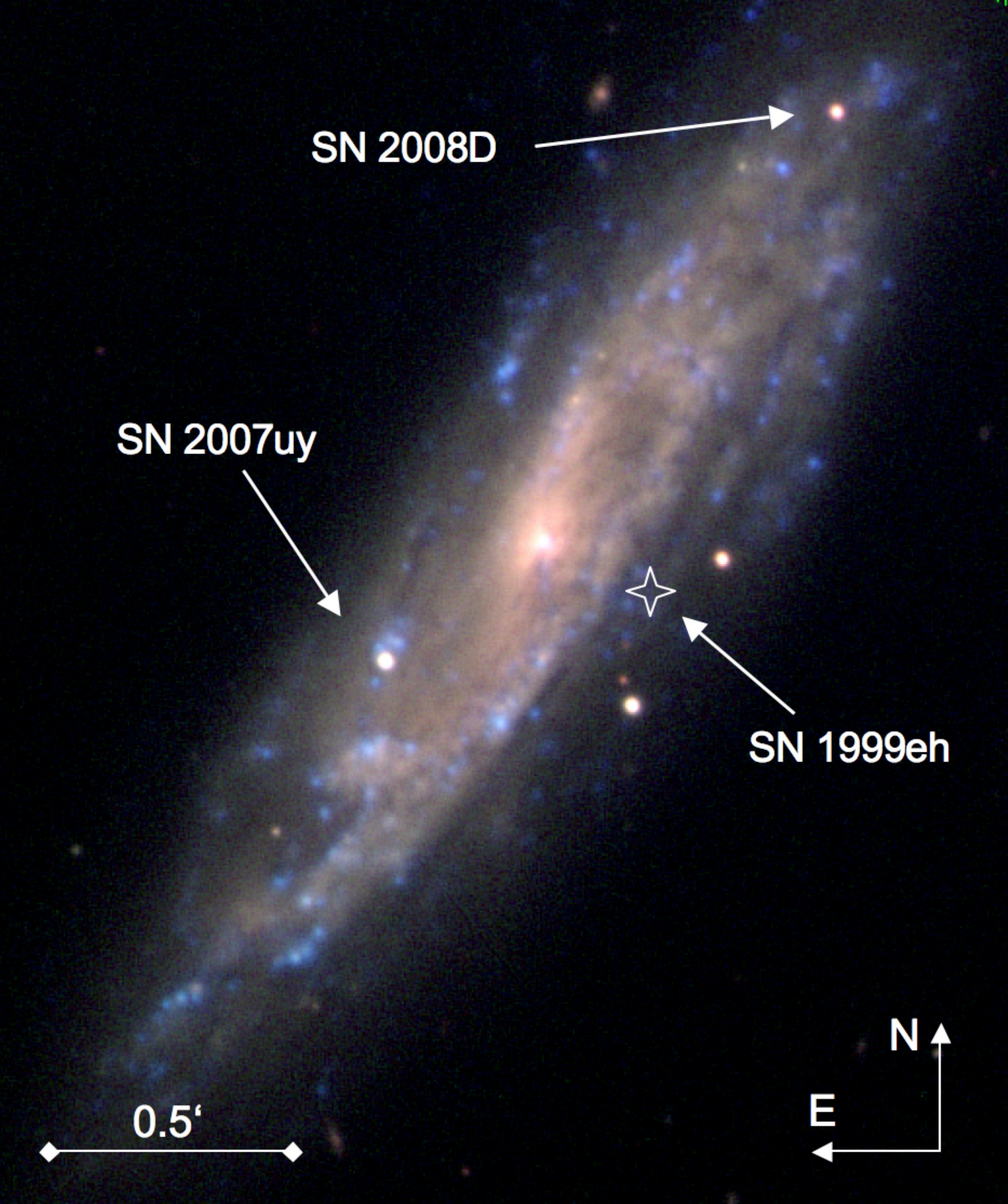}
\caption{Image of NGC 2770 from I (red color), H$\alpha$ (blue) and H$\alpha$ offband (green) filters. The field of view is about 2$\times$2.5 arcmin. Blue indicates H$\alpha$ emission which nicely show the SF regions in the spiral arms of the galaxy. The positions of the 3 SNe are shown of which two are still visible in the image.  
\label{NGC2770:radio}}
\end{figure}

For a log of the observations see Table \ref{NGC2770:obslog}.  Fig. \ref{NGC2770:radio} shows a color composite of I band observations taken with FORS1 \citep{Malesani08} and the H$\alpha$ and H$\alpha$ offband images used in this paper.

\begin{deluxetable}{lllll} 
\tablewidth{0pt} 
\tablecaption{Observation log}
\tablehead{\colhead{Date}&\colhead{telescope/instr.}&\colhead{grism/filter}&\colhead{$\lambda$, $\lambda_c$/$\Delta \lambda$}&\colhead{exptime}\\
\colhead{[UT]}&\colhead{}&\colhead{}&\colhead{$[$\AA$]$}&\colhead{[s]}}
\startdata
Jan.~11.31&VLT/FORS2&grism 300V&3000--9600&600\\
Jan.~13.07&NOT/ALFOSC&grism 4&2950--9050&3600\\
Jan.~15.21&NOT/ALFOSC&grism 4&2950--9050&600\\
Jan.~15.95&NOT/ALFOSC&grism 4&2950--9050&3600\\
Jan.~18.22&VLT/UVES&DIC1&3300--6650&3600\\
Feb.~03.15&NOT/ALFOSC&grism 4&2950--9050&1800\\
Mar.~16.11&VLT/FORS1&H$\alpha$ z=0.007&6604 / 64&120\\
Mar.~16.11&VLT/FORS1&H$\alpha$ z=0&6563 / 61&120
\enddata
\tablecomments{Log of the spectroscopic and narrowband imaging observations. \label{NGC2770:obslog}}
\end{deluxetable}

\section{Global properties of NGC 2770}\label{NGC2770:global}

\subsection{Modelling of the spectral energy distribution}
Extensive data are available in the literature on the emission of NGC 2770 at various wavelengths. We  performed photometry on the archival {\it GALEX} \citep{Martin03,galex2} UV images and Sloan Digital Sky Survey (SDSS) {\it ugriz} images. Furthermore, we obtained the infrared and radio fluxes from \citet{Cutri03, Moshir90, dressel78} and \citet{Condon98} reported in the NASA Extragalactic Database (NED). These data were used to fit the broad band spectral energy distribution (SED) with templates from the GRASIL code \citep{Silva98}.

In the GRASIL code each SED template is calculated by the following procedure. At the first stage an initial gas reservoir, infalling gas rate and star formation history are assumed and then at given time the emission of the resulting stellar population is summed up. Finally,  the total galaxy spectrum is calculated  by means of a two-dimensional radiative transfer method, applied to photons reprocessed by dust. After we found the best-fitting template we derived several galaxy properties from the SED \citep[as in][]{Michalowski08}. SFR, SN rate, stellar, dust, gas and total baryonic masses are given as output from GRASIL for the best-fiting template. The Infrared luminosity was obtained by integrating  the SED over the range of $8-1000\,\mu$m. The average extinction (outside of the molecular clouds) was calculated as: $A_V=2.5\log$ ($V$-band starlight extinguished by molecular clouds only / $V$-band starlight observed, see \citet{Silva98}). This parameter describes the extinction averaged throughout the galaxy as opposed to the line-of-sight extinction derived from optical GRB afterglows. $R_V$ was calculated comparing the extinction in the $V$- and $B$-bands: $R_V=A_V/E(B-V)$.

\begin{figure}
\includegraphics[width=\columnwidth]{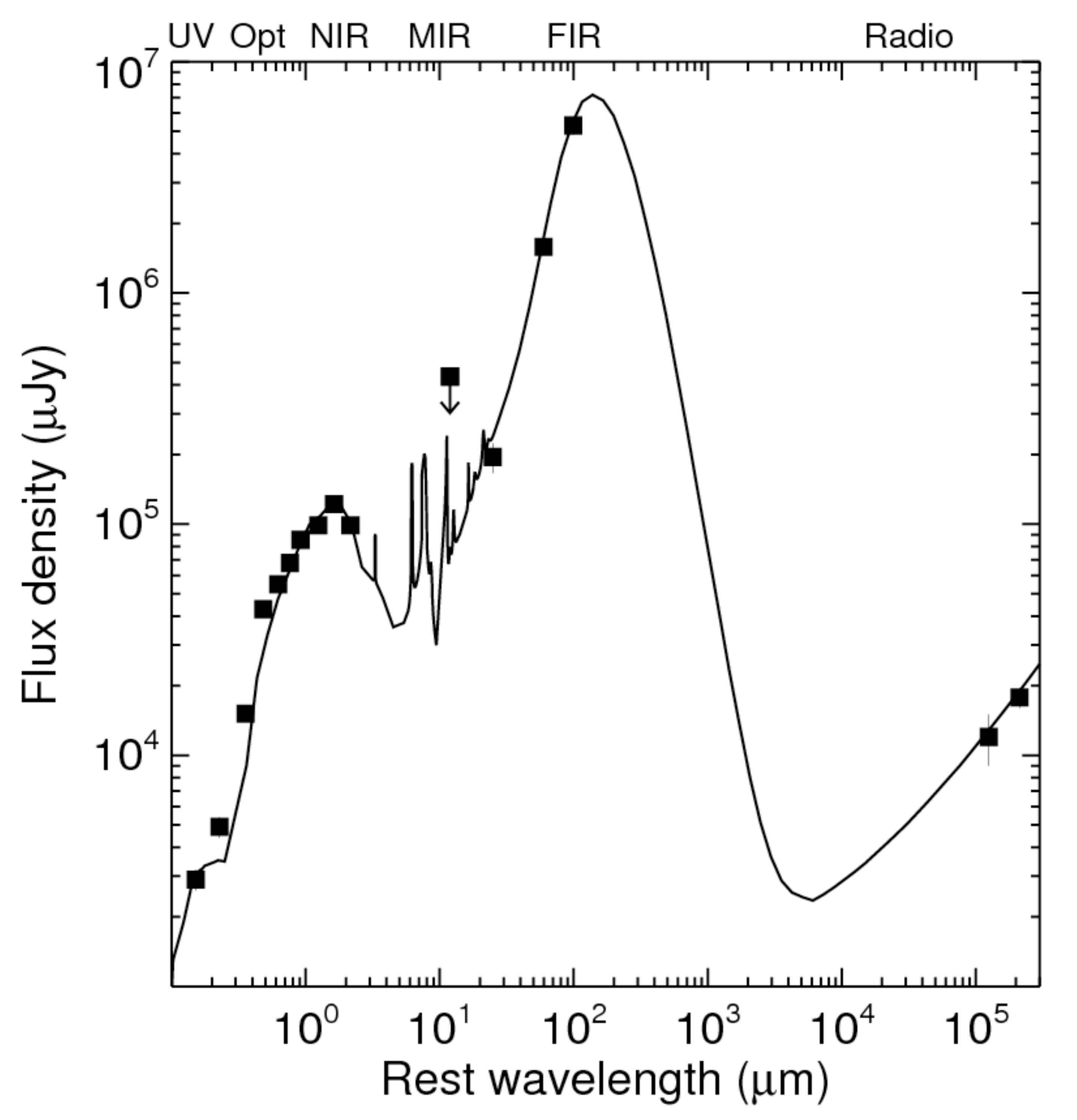}
\caption{Modelling of the broad-band SED of NGC 2770 from UV to radio wavelength using archival data compiled from the NASA Extragalactic Database (NED) and publicly available optical and ultraviolet data from the SDSS and GALEX archives. Error bars are mostly smaller than the symbols. The solid line shows the best-fit model from the GRASIL code \citep{Silva98}, which is the average of Sb and Sc templates (based on NGC 6946).
\label{NGC2770:SED}}
\end{figure}

The spectral energy distribution (SED) of NGC 2770 is well represented by an average model composed of the spiral Sc (NGC 6946) and Sb galaxies, taken from \citet{Silva98}. From the SED fit (see Fig.~\ref{NGC2770:SED}) we derive a star-formation rate (SFR) of $\sim1.1$M$_\odot$ yr$^{-1}$ and a stellar mass of 2.1$\times$10$^{10}$ M$_\odot$ \citep[see][for details on how these are derived from the SEDs]{Michalowski08}. The resulting specific star-formation rate (SSFR) is 0.05 Gyr$^{-1}$. The SFRs derived from different methods (radio, IR, UV, SED modelling) differ by about a factor of 2. This is not surprising since the UV SFR estimate is affected by extinction and the different methods trace different parts of SF in the galaxy. Still, the values indicate that the SFR is not particularly high in NGC 2770. The SFR in radio is also directly connected to the supernova rate (SNR) as the nonthermal radio flux is produced by relativistic electrons accelerated in the SN shocks. Also the supernova rate (SNR) both from radio data and the SED modelling is therefore rather low with 0.01--0.02 SNe yr$^{-1}$ which is comparable to the SNR of the Milky Way (MW). In addition to the SED fit, we also searched the literature for global measurements and properties of NGC 2770. Those, together with the SED outputs, are presented in Table \ref{NGC2770:globalprop}.

\subsection{NGC 2770 in the context of other spiral galaxies}

In order to quantify to what extent the properties of NGC 2770 are similar to other galaxies we 
compare the global properties of NGC 2770 with a sample of more than 100 nearby spiral galaxies 
from \cite{BroeilsRhee} which also includes NGC 2770. This sample is based on two HI surveys of spiral and irregular galaxies with the Westerbork Synthesis Radio Telescope. We then determine the median and mean values of a number of properties and compare that with the values derived for NGC 2770. The mean in this sample is dominated by a few extreme outliers in the sample, therefore, we prefer to take the median value for comparison. We then conclude the following:

\begin{itemize}
\item The HI surface density is one of the highest in the sample (median: 3.65 M$_\odot$/pc$^2$, maximum: 8.05 M$_\odot$/pc$^2$, NGC 2770: 4.99 M$_\odot$/pc$^2$).
\item The HI mass is also above average (median: 3.5 $\times$10$^9$ M$_\odot$, NGC 2770: 7$\times$10$^9$ M$_\odot$ ).
\item The total mass is above the median but slightly below the mean, however this is dominated by some very massive galaxies (median: log M$_\mathrm{tot}$=7.8 M$_\odot$, NGC 2770: log M$_\mathrm{tot}$=11.38 M$_\odot$).
\item The ratio of the gas mass to the blue ($B$-band) luminosity is close to average (median: 0.26 M$_\odot$/L$_B\odot$, NGC 2770: 0.24 M$_\odot$/L$_B\odot$).
\item The ratio of the gas mass to the near-IR ($H$-band) luminosity is slightly above average (median: 0.31 M$_\odot$/L$_H\odot$, NGC 2770: 0.44 M$_\odot$/L$_H\odot$).
\item The ratio of the gas mass to the total mass is slightly above average (median: 0.05, NGC 2770: 0.06).
\end{itemize}

This comparison shows that NGC 2770 has more or less average properties within this sample, except for the HI mass. The rather high gas mass and HI surface density indicate that there is a lot of material present to produce stars. However, the current and past SFRs as derived from different datasets (see also Section \ref{NGC2770:Halpha}) are not at all exceptionally high. \cite{Soderberg08} suggested that the companion galaxy NGC 2770B could be interacting with NGC 2770 and cause the enhancement in SF. No obvious perturbations are visible in the optical images of NGC 2770. It would be interesting to investigate possible perturbations in the HI velocity field, but the available data from VLA do not provide a good enough resolution. \cite{Wainwright07} conclude from {\it HST} imaging of GRB host galaxies that about 30--50\% show evidence for interaction. However, the host of GRB\,980425/SN 1998bw was found to be an isolated galaxy \citep{Foley06} and the SF history is also consistent with a constant SFR over a few Gyrs \citep{Sollerman05}.

\begin{deluxetable*}{llll} 
\tablewidth{0pt} 
\tablecaption{Global properties of NGC 2770}
\tablehead{\colhead{Property} & \colhead{Value} & \colhead{Notes}& \colhead{Reference}}
\startdata
SFR (SED)&1.1 M$_{\odot}$ yr$^{-1}$& 	&this work\\
SFR (radio)&0.6 M$_{\odot}$ yr$^{-1}$& using \citet{YunCarilli02} &this work\\
LIR&1.4$\times$10$^{10}$ L$_{\odot}$& &this work\\
SFR (IR) &1.22 M$_{\odot}$ yr$^{-1}$&using \citet[][eq. 4]{Kennicutt92}  &this work\\
SFR (UV) & 0.50 M$_{\odot}$ yr$^{-1}$&using \citet[][eq. 1]{Kennicutt92}& this work\\
v$_\mathrm{rot}^\mathrm{max}$ & 152 km s$^{-1}$ & 21cm emission&\citep{RheeAlbada}\\
M$_\mathrm{gas}$&1.9$\times$10$^9$ M$_{\odot}$&&this work\\
N$_\mathrm{HI}$ & 1.6$\times$10$^{20}$ atom cm$^{-2}$&&\citep{Staveley-Smith88} \\
M$_\mathrm{HI}$ & 7$\times$10$^9$ M$_{\odot}$ &&\citep{BroeilsRhee}\\
M$_\mathrm{star}$& 2.1$\times$10$^{10}$ M$_{\odot}$&SED&this work\\
M$_\mathrm{tot}$& 2.6$\times$10$^{10}$ M$_{\odot}$&SED& this work\\
M$_\mathrm{tot}$ &1.1$\times$10$^{11}$ M$_{\odot}$&including DM; from v$_\mathrm{rot}$ &\citep{BroeilsRhee}\\
M$_\mathrm{dust}$& 1.9$\times$10$^7$ M$_{\odot}$&SED&this work\\
M$_\mathrm{BH}$& 1.3$\times$10$^6$ M$_{\odot}$&&\citep{DongRobertis06}\\
SN rate SED& 0.02 SN yr$^{-1}$ &&this work\\
SN rate radio&0.01 SN yr$^{-1}$&using \citet[eq. 18 in][]{Condon92}&this work\\
A$_\mathrm{V}$& 0.15 mag&SED&this work\\
A$_\mathrm{B}$& 0.19 mag&SED&this work\\
A$_\mathrm{B}$ & 0.14 mag & from N$_{HI}$ &\citep{Staveley-Smith88} \\
E(B--V)& 0.03 mag &SED&this work\\
E(B--V) & 0.84 mag &Balmer decrement, global& \citep{Ho}\\
R$_\mathrm{V}$& 4.3&SED&this work\\
n$_\mathrm{e}$ & 54 cm$^{-3}$&(electron density)&\citep{Ho}\\
\enddata
\label{NGC2770:globalprop}
\end{deluxetable*}

\section{Spatially resolved optical spectroscopy of NGC 2770}\label{NGC2770:parts}

From the four 2-dimensional longslit spectra, we extracted traces at all positions along the slit that showed emission in H$\alpha$.  We chose equally large extraction windows in all the spectra with a width of 3 arcsec, corresponding to 0.43 kpc. The spectra were then extracted using the strong
and well defined SN traces as templates. For the slit containing SNe 2008D and 2007uy we took the SN trace closest to the region we wanted to extract. The relative distortions along the spatial axis are
however only within one to two pixels and even for spectra with only one template trace, we are confident that we extracted the same region along the entire dispersion axis. We note here that the FORS spectrograph has an atmospheric dispersion corrector while ALFOSC has none. This implies that the ALFOSC slit do not necessarily trace the same physical region in the blue and in the red part of the spectra. However, the airmass at the time of the observations was very low and the expected refraction should be negligible. 

For the wavelength calibration, we extracted several traces from the arc spectrum with the SN trace as a template and shifted to different positions along the spatial axis. In each of these traces we identified the arc lines and wavelength calibrated the different parts along the slit with the nearest arc spectrum trace. Fortunately, the distortion of the arclines along the entire chip is also within a few pixels. The maximum deviation between the arclines at the extracted trace and the position taken for the arcspectrum should be less than 1 pixel which corresponds to an accuracy of 0.3 \AA{} comparable to the accuracy from the wavelength calibration itself. Finally, we flux calibrated the individual spectra using the corresponding standard stars observed at the same night and the same setting. We note that the flux of an extraction window is different from the one of a pointsource as the point spread functions from the neighbouring pixels in the spatial direction extend into the extracted part. 

\begin{figure*}
\includegraphics[width=8.5cm]{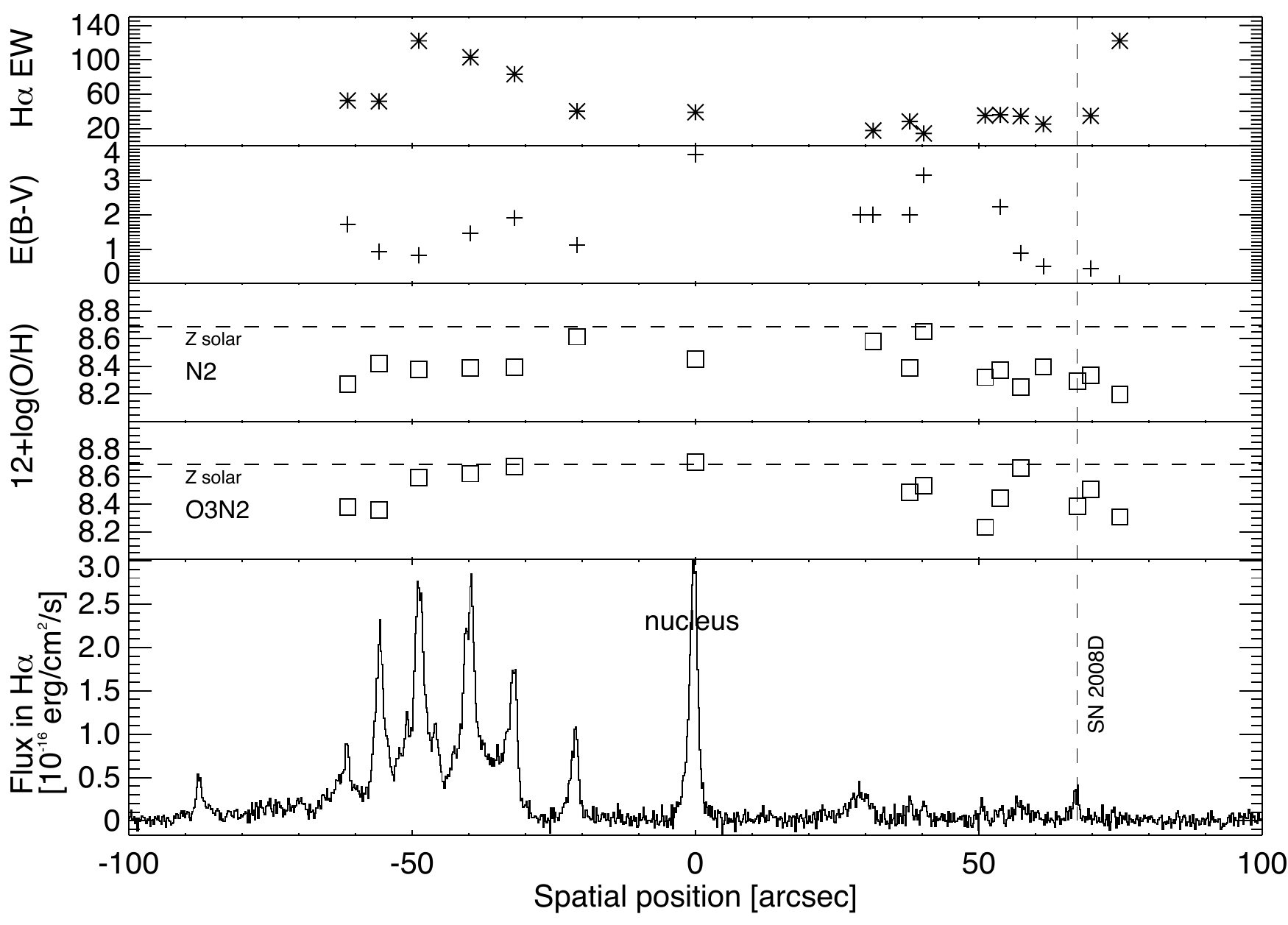}
\includegraphics[width=8.5cm]{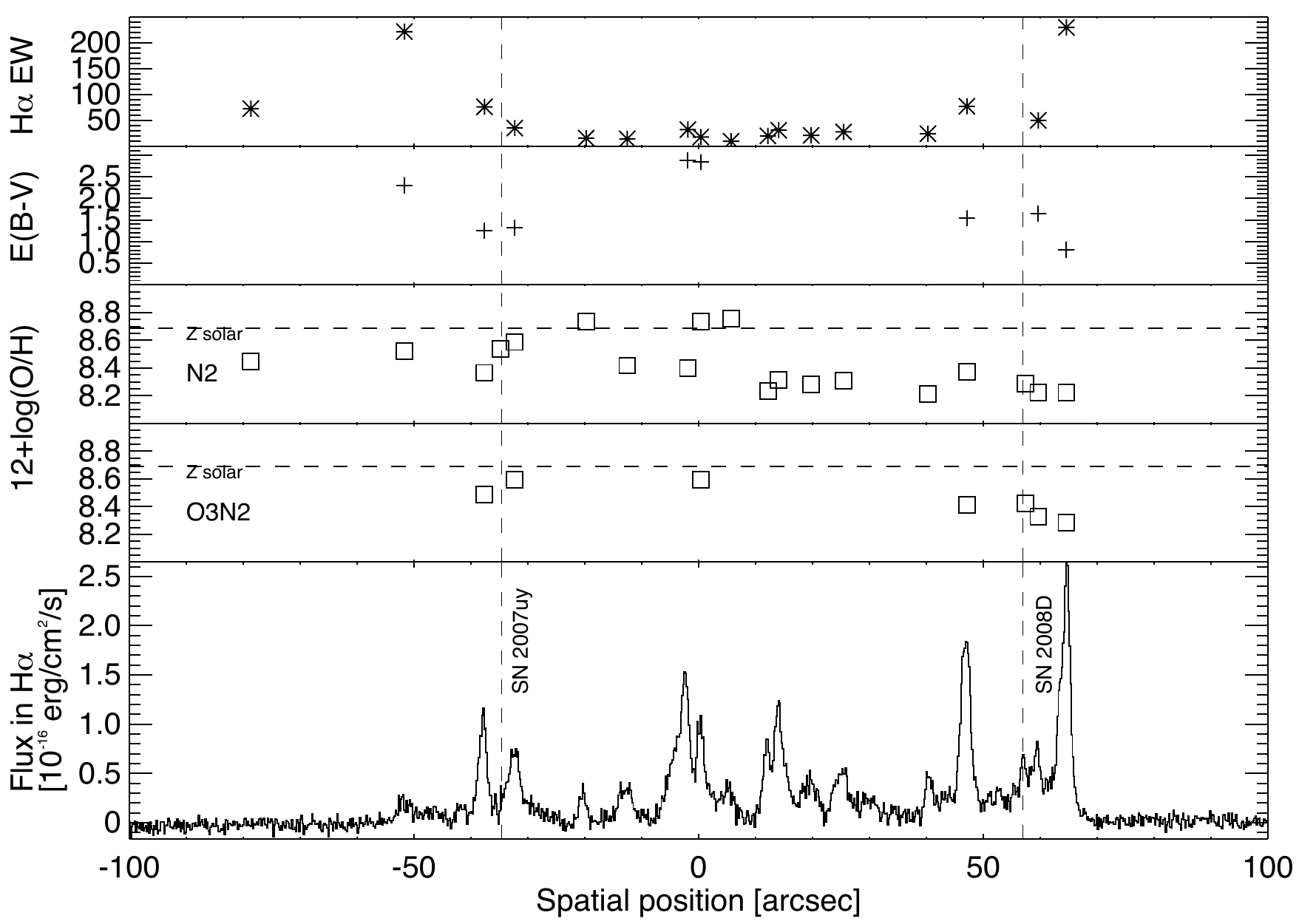} \\
\includegraphics[width=8.5cm]{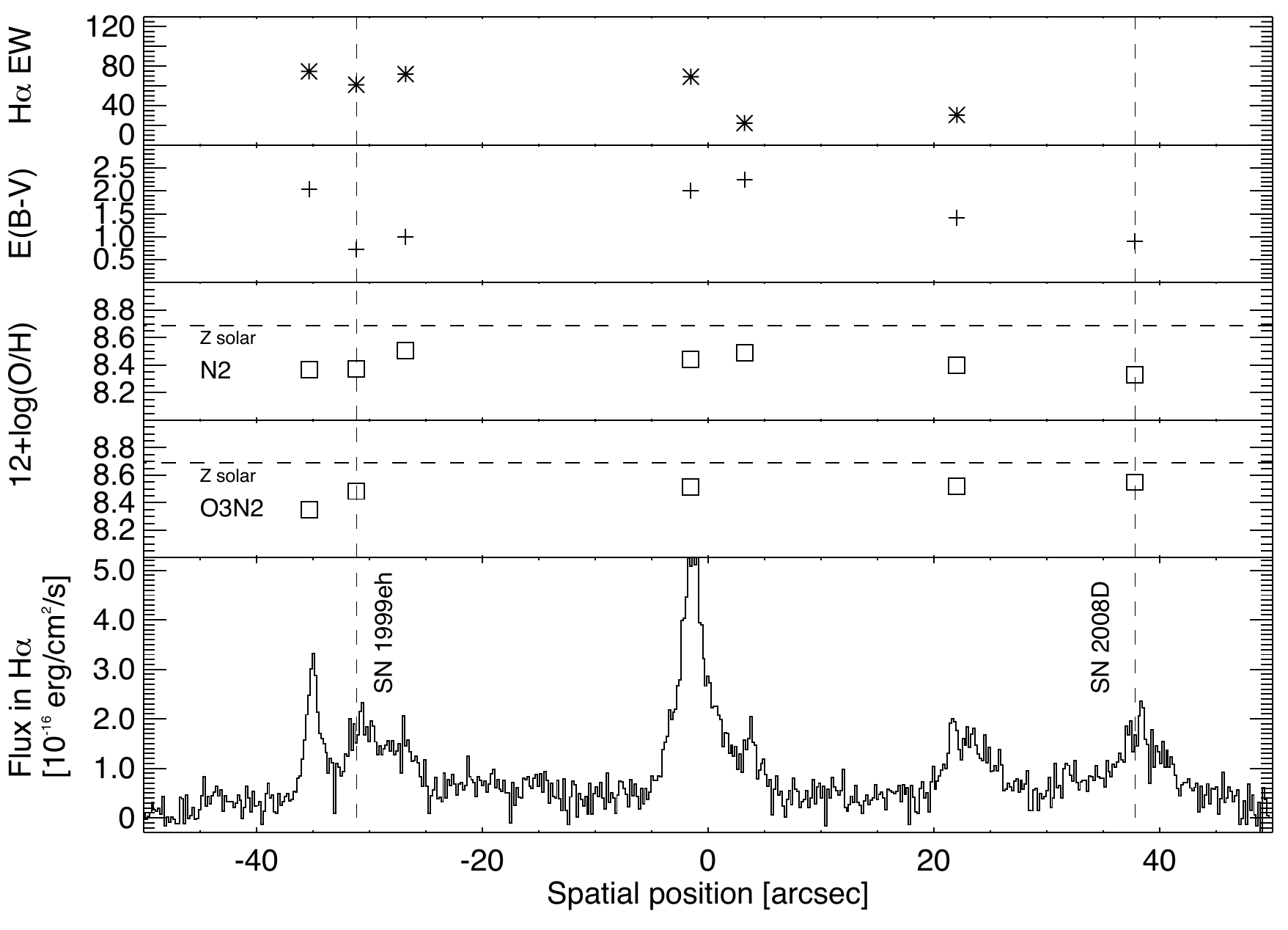}
\includegraphics[width=8.5cm]{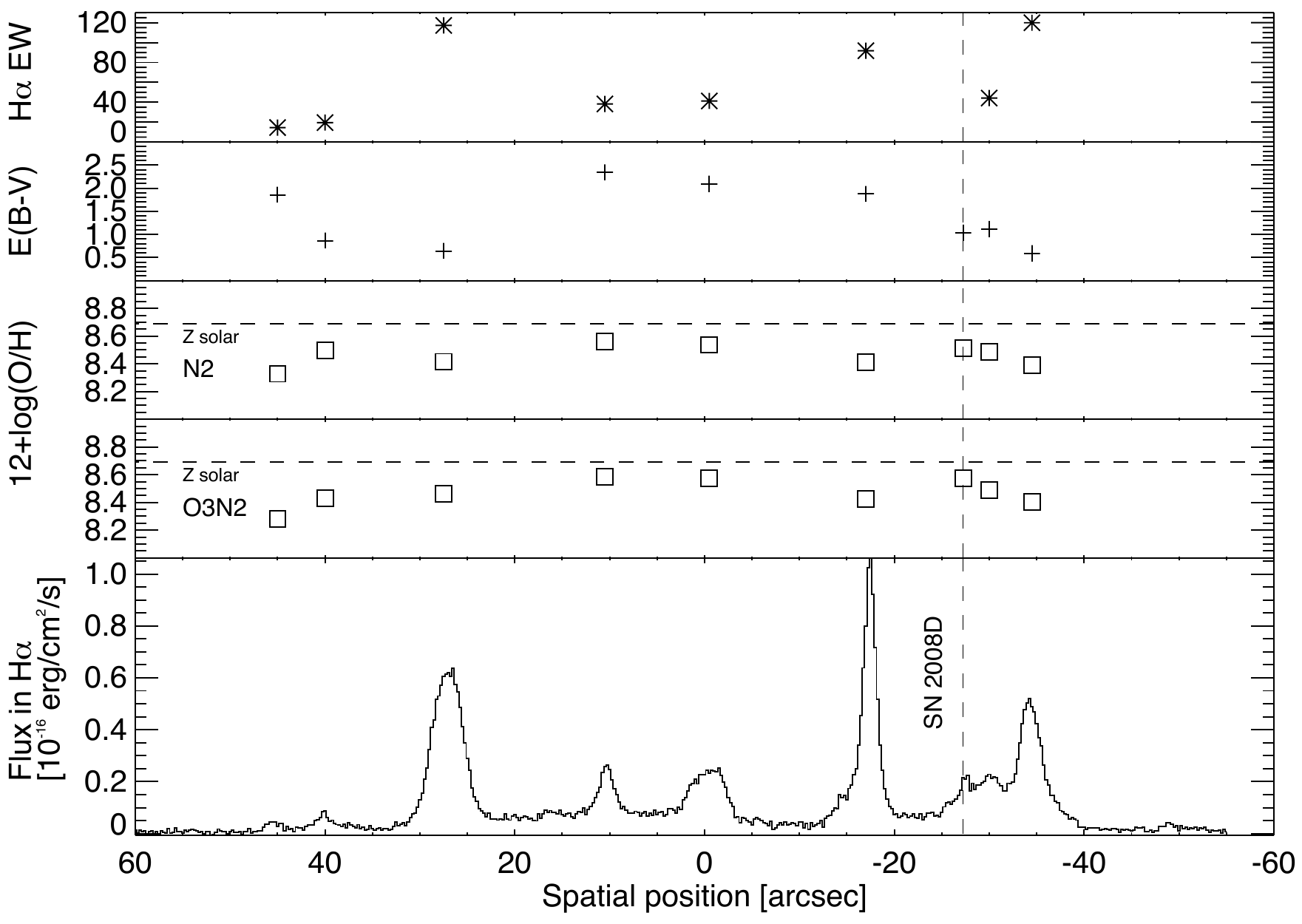}
\caption{Cut through the H$\alpha$ line and the properties at the different HII regions (see Fig. \ref{NGC2770:slitmetallicities} for the slitpositions on the galaxy). We plot two calibrations for the metallicity (errors are 0.2 dex for both methods), the extinction E(B--V) in [mag] and the H$\alpha$ EW in [\AA]. The four panels are: upper left: ALFOSC slitposition covering the major axis of the galaxy (slit1), upper right: the slit going through the sites of SN 2008D and SN 2007uy (slit2), lower left: position including SN 2008D and 1999eh (slit3) and lower right: the FORS slitposition (slit4). For the slit along the major axis, spatial position = 0 kpc indicates the center of the galaxy, for the other positions, for the other slits, 0 arcsec was chosen randomly.
 \label{NGC2770:slitplots}}
\end{figure*}

We then fitted the emission lines present in the spectra with Gaussians and measured the fluxes and, where possible, the equivalent widths (EW) of the H$\alpha$ line. For the SN regions, the latter is prevented by the contamination from the SN continuum. In a few cases, deblending the [N II] and H$\alpha$ lines in the ALFOSC spectra proved to be difficult at this low resolution. In some regions, it seems that the part covered by the slit actually contains two kinematically different regions as the H$\alpha$ lines shows a double peak which then blends with the bluer component of the corresponding [N\,{\sc ii}] double
peak. This is not surprising since we observe NGC 2770 at a small inclination. Regions with heavily blended lines were discarded from the analysis. The properties in the different regions derived from the emission line analysis as present in the next sections are listed in Tab. \ref{NGC2770:local} and plotted in Fig.~\ref{NGC2770:slitplots}.

\begin{deluxetable}{lllcc} 
\tablewidth{0pt} 
\tablecaption{Resolved properties of NGC 2770}
\tablehead{\colhead{Position} & \colhead{12+log(O/H)} & \colhead{E(B-V)} & \colhead{H$\alpha$ EW} & \colhead{SFR}\\
{[arcsec]}&\colhead{(O3N2)~~(N2)}&\colhead{[mag]}&\colhead{[\AA]}&\colhead{\tiny [10$^{-3}$M$_\odot$/yr/kpc$^2$]}}
\startdata
    74.86		&	8.31  ~~~~~  8.20		& 0.01	& 122.1 & 5.2	\\
    69.73   		&	8.50  ~~~~~ 8.33		& 0.44	& 34.70& 4.9	\\
    67.45 {\tiny (08D)}  &	8.38  ~~~~~	8.30		& ---	& ---	&2.7	\\
    61.37 		&	---	~~~~~~~~ 8.40	&   ---& 25.04& 2.1	\\
    57.38  		&	8.66  ~~~~~ 8.25		&0.89	&34.32 & 5.1	\\
    53.77		&	8.44  ~~~~~ 8.37		& 2.22	&35.75	& 5.3	\\
    51.11 		&	8.23   ~~~~~ 8.32		& 4.20	&35.09	& 4.9	\\
    40.28		&   	8.53   ~~~~~ 8.65		& 3.12	&14.21	& 3.1	\\
    37.81		&	8.49   ~~~~~ 8.40		& 2.00	&28.19	& 6.7	\\
    31.35   		&	---  ~~~~~~~~ 	8.58		&    ---&17.55& 3.6		\\
     0.0   	 	& 	8.71	  ~~~~~  8.45		& 3.74	&38.85	& 38.2	\\
     --20.90    	&	--- ~~~~~~~~ 8.61		& 1.12	&40.08	& 4.0	\\
     --31.92		&   	8.67	~~~~~	8.40 		& 1.90	&83.27	& 14.7\\
     --39.71		&	8.62   ~~~~~	8.39		& 1.62	&102.9	& 15.6	\\
     --48.83		&	8.59   ~~~~~	8.38		& 0.82	&122.1& 18.2	\\
     --55.86		&	8.36  ~~~~~	8.41		& 0.93	&51.63	& 10.2	\\
     --51.37		&	8.38  ~~~~~	8.28		& 1.72	&52.52	& 5.8\\ \hline
     64.60		&	8.28   ~~~~~ 	8.22		& 0.82	&229.8& 15.8\\
     59.66    	&	8.32  ~~~~~  	8.23		&1.64	&50.07 & 4.9\\
     57.38 {\tiny (08D)} 	&8.42 ~~~~~     8.29	  	& ---	& --- & 3.9\\
     47.12		&	8.41  ~~~~~    8.37		& 1.27	&77.29 &9.7\\
     40.28		&	--- ~~~~~~~~  8.21		&  ---	&24.31& 3.1\\
     25.46		&	---  ~~~~~~~~ 8.31		&  ---	&27.85& 4.9\\
     19.76		&	---  ~~~~~~~~ 8.29		& ---	&21.16& 4.4\\
     14.06   		&	--- ~~~~~~~~ 8.31		& ---	&31.13& 8.0\\
     12.16   		&	--- ~~~~~~~~  8.24		& ---	&20.14& 4.5\\
     5.70    		&	--- ~~~~~~~~  8.76		&  ---	&9.743& 3.7\\
     0.38     		&	8.60  ~~~~~    8.74		& 2.91	&17.89& 7.3\\
     --1.90    	&	--- ~~~~~~~~	  8.40	& 2.88	&32.19& 12.0\\
     --12.54    	&	--- ~~~~~~~~	 8.42		&  ---	&14.37& 3.9\\	
     --19.76 	&	--- ~~~~~~~~	  8.73	&  ---	&15.95& 2.8\\
     --32.30		&	8.59   ~~~~~    8.59		& 1.41	&35.29& 10.9\\
     --34.77{\tiny (07uy)}&  --- ~~~~~~~~	  8.53		& ---	&---	& 8.1\\ 
     --37.62		 &	8.49   ~~~~~   8.37	& 1.37	&76.20& 15.1\\
     --51.68		&	--- ~~~~~~~~ 8.52		& 2.23	& 221.6& 3.6\\
     --78.66		&	--- ~~~~~~~~	  8.45		&   ---& 72.64 & 3.5\\ \hline
         37.81 {\tiny (08D)}	&	---  ~~~~~~~~   8.33	& 	---	&	---& 10.7\\
     22.04		&	8.52 ~~~~~   	8.40			&	1.41 	&	30.53& 10.2\\
     3.23		&	---	 ~~~~~~~~	8.49			&	2.25	&	22.36& 9.8\\
     --1.52		&	8.51 ~~~~~   8.44			&	2.00	&	69.33& 32.9\\
     --26.79		&	--- ~~~~~~~~  8.51		&	---&	71.93& 11.1\\
     --31.16 {\tiny (99eh)}&	8.48 ~~~~~   8.37	&	0.72	&	61.23& 11.8\\
     --35.34		&	  8.35  ~~~~~    8.37		&	 2.04	&	74.66 & 11.2\\ \hline
     34.50		&	8.40 ~~~~~	 8.39		&	0.59	&	119.9& 2.8\\
     30.00		&	8.49  ~~~~~  8.48		&      1.12	&	44.25& 1.2\\
     27.25 {\tiny (08D)}	&	8.57 ~~~~~     8.51		&    0.95	&	5.014& 1.1\\ 
     17.00		&	8.43  ~~~~~  8.41		&      1.88	&	 91.78&4.1\\
     0.500		&	8.57  ~~~~~ 8.54		&      2.10	&	41.34&1.4\\
     --10.50		&	8.58  ~~~~~  8.56		&      2.34	&	38.31&1.1\\
     --27.50		&	8.46  ~~~~~  8.41		&     0.64	&	117.2& 3.7\\	
     --40.00		&	8.43  ~~~~~  8.50		&   0.87	&	19.61& 0.3\\
     --45.00		&	8.28  ~~~~~  8.33		&      1.86	&	14.56& 0.2\\ 
    \enddata
\tablecomments{Measurements from the optical spectra, from top to bottom: slit through the major axis (slit1), slit through SN 2007uy and SN 2008D (slit2), slit through SN 2008D and SN 1999eh (slit3) and FORS spectra through SN 2008D (slit4). Metallicities and SFRs were calculated from the fluxes not corrected for the extinction. The error in the metallicities are around 0.2 dex. \label{NGC2770:local}}
\end{deluxetable}

\subsection{Metallicity}\label{metallicities}

For the different regions along the slits we estimated the metallicity using the N2 parameter.
This is the ratio between [N\,{\sc ii}]/H$\alpha$ and is  calibrated as 12+log(O/H)=9.37 + $2.03\times$N2 + $1.26\times$N2$^2$ + $0.32\times$N2$^3$ \citep{Pettini04}. In the regions with significant detection of [O\,{\sc iii}], we also use the O3N2 parameter, $O3N2$=log[(O\,{\sc iii} $\lambda$ 5007 /H$\beta$)/(N\,{\sc ii}/H$\alpha$)], to derive metallicities with 12+log(O/H)=8.73~--~0.32\,O3N2 \citep{Pettini04}.  

The O3N2 parameter has been calibrated against oxygen abundances derived directly 
by determining the electron temperature T$_e$ from the temperature sensitive [O\,{\sc iii}] $\lambda$ 4363 line \citep[e.g.][]{Izotov06}. The O3N2 parameter is also considered as the most
reliable method by \cite{KD02}, if the spectra could be corrected for dust extinction. 
However, for low metallicities, the [O\,{\sc iii}] $\lambda$ 5007 line becomes very weak. We were not able to use it for deriving metallicities everywhere, since in some parts of the spectra it was only detected with very low significance. For consistency, we therefore use the N2 estimate throughout this article.
The N2 parameter does, however, saturate at about solar metallicity and even turns around at higher metallicities (about twice solar). This is not a problem in our data as we do not have regions where the O3N2 parameter gives supersolar metallicities while the N2 metallicity gives a subsolar value.  

\begin{figure*}
\centering
\includegraphics[width=15cm]{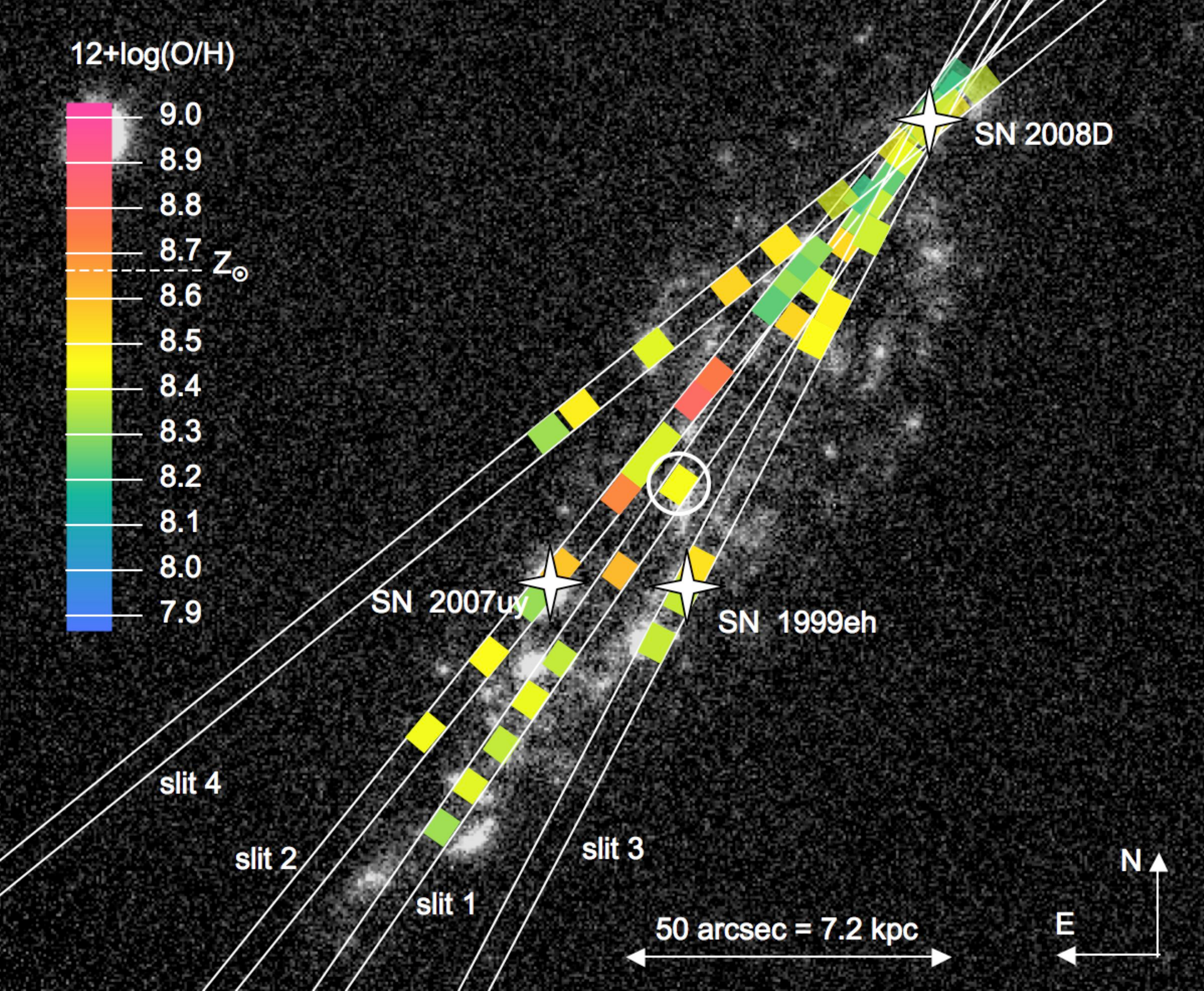}
\caption{Metallicities in NGC 2770 along the 4 slitpositions as described in Fig. \ref{NGC2770:slitplots} including the 3 SN sites derived from the [N\,{\sc ii}]/H$\alpha$ ratio. The image of the galaxy was taken with the H$\alpha$ filter in order to highlight the starforming regions. The circle indicates the center of the galaxy.
\label{NGC2770:slitmetallicities}}
\end{figure*}

We find that the N2 and the O3N2 methods give rather consistent results at the sites where both are determined. The error of the O3N2 parameter is about 0.2 dex in the ALFOSC spectra and 0.10--0.15 in the FORS slit. The metallicities from the O3N2 and N2 parameters have systematic errors of 0.14 dex for O3N2 and 0.18 for N2. We therefore estimate the total error of each metallicity value to be 0.2 dex for both O3N2 and N2. In Fig.~\ref{NGC2770:slitplots} we plot the results from both methods where available. Figure~\ref{NGC2770:slitmetallicities} visualizes the 2D distribution of the N2 metallicities in the different parts of the galaxy as probed by the four slits. 

The metallicity at the SN sites derived from N2 are 12+log(O/H) $\sim$
8.4, 8.5 and 8.4 for SNe 2008D, 2007uy and 1999eh respectively,
which corresponds to 0.55--0.70 Z$_\odot$ [when using 8.66 for the Sun
\cite{Asplund04}]. The metallicities derived at the position of SN 2008D differ between the ALFOSC and the FORS slits, but still agree within the errors. We therefore adopt an average metallicity of 8.4 for SN 2008D. The other regions in the outer spiral arms on both sides of the galaxy give similar values for the metalicity for different slits.  The nucleus and the central regions have solar or supersolar metallicity. From this we derive a metallicity gradient of $-0.06$ dex per kpc. The average metallicity computed from the values along the major axis of the galaxy is 8.4. 

It is interesting to note that the metallicities at the SN Ib sites in NGC 2770 lie between the low values found for GRB connected broadline SNe Ic and those of broadline SNe Ic not connected to GRBs \citep{Modjaz08a}. They are also slightly higher than the metallicity of the HII region which hosted SN 1998bw, a broadlined SN Ic  connected to GRB\,980425, as determined from IFU observations of the host \citep{Christensen08}. Also the site of the SN-less GRB\,060505 \citep{Thoene08} had  a lower metallicity as the SN Ib sites in NGC 2770. It seems, GRB related SNe have a rather low metallicity while SNe Ib have about half solar to solar metallicities whereas (broadline) SNe Ic require solar to slightly above solar metallicities. This result seems consistent with the suggestion that in nonrotating stars, the mass loss scales with the metallicity \citep[e.g][]{Crowther02} and SN Ic progenitors are likely experiencing larger mass loss than SNe Ib. Fast rotating stars, however, as required by the collapsar model \citep{MacFadyen99}, should preferentially have low metallicities and less mass loss in order to keep the high angular momentum, which is required to produce a GRB.

\subsection{Extinction}
The extinction throughout the galaxy is estimated from the Balmer decrement. Since H$\gamma$ is not detected, we can only use the H$\alpha$/H$\beta$ ratio \citep{Osterbrock89}. The derived extinction is
generally high with values for E(B--V) between 0.4 and 3.7 mag with most of the regions having E(B--V) $\sim$ 1 mag. The extinction at the SN sites is considerable with E(B--V)$=$0.9 mag for
SN 2008D, 1.4 mag for SN 2007uy and 0.7 mag for SN 1999eh. \cite{Malesani08} get a value of E(B--V)$=$0.8 mag from fitting the SN SED, while \citep{Modjaz08b} adopt 0.6$\pm$0.1 from the SN broadband fit. The Galactic extinction towards NGC 2770 is negligible with E(B--V)$=$ 0.02 mag \citep{Schlegel98}.

\begin{deluxetable}{llllll} 
\tablewidth{0pt} 
\tablecaption{EWs and column densities from the UVES spectra}
\tablehead{\colhead{line id}&\colhead{$\lambda_\mathrm{rest}$}&\colhead{$\lambda_\mathrm{obs}$}&\colhead{EW}&\colhead{b}&\colhead{log N}\\
\colhead{}&\colhead{$[$\AA$]$}&\colhead{$[$\AA$]$}&\colhead{$[$\AA$]$}&\colhead{$[$km\,s$^{-1}$$]$}&\colhead{$[$cm$^{-2}]$}}
\startdata
Ca II K &	3933.66	& 3961.00&	0.55$\pm$0.03 & 14.1$\pm$7.1 & 12.92$\pm$0.07\\
	&			& 3961.20	&				& 18.3$\pm$2.4 & 12.78$\pm$0.07\\
Ca II H &	3968.47	& 3996.05	&	0.42$\pm$0.03	& 14.1$\pm$7.1 & 12.92$\pm$0.07\\
	&			& 3996.25	&				& 18.3$\pm$2.4 & 12.78$\pm$0.07\\
Na I D2 &	5889.95	& 5930.92	&	0.33$\pm$0.02	& 7.3$\pm$0.3 & 12.96$\pm$0.05 \\
		&		& 5931.21	&	0.38$\pm$0.02	& 8.1$\pm$0.3 & 12.71$\pm$0.02\\ 
Na I D1 &	5895.92	& 5936.94	&	0.26$\pm$0.02 & 7.3$\pm$0.3 & 12.96$\pm$0.05 \\
		&		& 5937.23&	0.36$\pm$0.02	& 8.1$\pm$0.3 & 12.71$\pm$0.02\\  \hline
DIB &	5780.37	& 5821.5	&	0.245$\pm$0.04& ---			& ---\\
DIB &	5796.99	& 5837.4	&	0.850$\pm$0.05& ---			& ---\\
DIB &	6283.85	& 6328.0	&	0.090$\pm$0.02& ---			& ---
\enddata
\label{NGC2770:UVES}
\tablecomments{Equivalent width measurements, b parameters and column densities (where applicable) of interstellar absorption lines in NGC 2770 along the line of sight to SN 2008D, all wavelengths are in air. EWs are given for both components in the Ca II and Na I absorption systems. Galactic rest DIB wavelengths are from Galazutdinov et al. (2000)}
\end{deluxetable}

Further evidence for dust extinction along the line of sight towards SN 2008D comes from the detection of NaD absorption in the UVES high resolution spectra, the EWs measured are listed in Tab. \ref{NGC2770:UVES}. For unsaturated and mildly saturated lines, there is a direct relation between the strength of the Na D absorption and the extinction \citep{Munari97} which have been calibrated towards stars in the LMC. Taking the EW for the two components together, we get an extinction of E(B--V)$=$1.1 mag. However, \cite{Munari97} note that systems with such high EWs are usually consisting of several components and the extinction of different components is additive so it is more appropriate to add the extinction of the resolved subcomponents which are on the linear part of the EW--E(B--V) relation. The two components then give E(B--V)$_1$ = 0.14 and E(B--V)$_2 =$ 0.18 mag and therefore E(B--V)$_\mathrm{tot}$ = 0.32 mag. There is also a high N$_H$ column derived from the X-ray spectra \citep{Soderberg08} of N$_H$ = (7$\pm$1)$\times$10$^{21}$ cm$^{-2}$, assuming Galactic abundances. The corresponding gas-to-dust ratio of 2.8$\times$10$^{21}$ cm$^{-2}$mag$^{-1}$ is low (close to the Galactic ratio N$_H$ = 1.7 10$^{21}$ cm$^{-2}$mag$^{-1}$) compared to the ratios observed in GRB host galaxies that are typically an order of magnitude higher, suggesting that NGC 2770 is dust-rich compared to those hosts.

We also detect a number of diffuse interstellar bands (DIBs) at $\lambda\lambda$ 5821.5 \AA, 5837.4 \AA{} and 6328.0 \AA{} in the UVES spectra which indicates the presence of dust at the site of SN 2008D. DIBs are speculated to be produced by large carbon molecules, possibly related to the polycyclic aromatic hydrocarbons (PAHs) that produce strong fluorescence lines in the mid-infrared. It is interesting to note that DIBs have never been detected in the afterglow spectra of GRBs suggesting a harder radiation field and younger stars in those galaxies compared to NGC 2770. We fit the observed spectra (rebinned to 0.09\AA) with template DIB profiles obtained from high-resolution high-S/N spectra of DIBs in the diffuse cloud toward HD~144217. We keep the width of the feature fixed but leave the continuum, the central wavelength and the depth scaling as free parameters. Doppler broadening due to the different velocity components as fitted from the NaD and Ca II absorption in the UVES spectra (see Section \ref{NGC2770:kinematics}) is negligible for the broad $\lambda$ 6283 and $\lambda$ 5780 DIBs (FWHM $\sim$~100 and $\sim$~200 km\,s$^{-1}$). For the narrow $\lambda$ 5797 DIB (FWHM $\sim$~40 km\,s$^{-1}$) we assume that DIBs are present in similar abundance in both cloud components as is the case for the Ca II absorption strenghts. Taking into account the two components increases the derived $\lambda$ 5797 DIB strength by about 10-15\%.

Diffuse interstellar bands can provide an important measure, just as Na I and K I, for the amount of dust in line-of-sights probing the diffuse interstellar medium. Additionally, the behavior of different DIBs (or DIB ratios) in the Milky Way can be used to gain insight in the local physical conditions of extra-galactic diffuse clouds. The $\lambda$ 5780 DIB carrier is sensitive to UV radiation and becomes stronger when exposed to sufficient UV (at which time for example the $\lambda$ 5797 DIB carrier is already being destroyed by this radiation). For example, the DIB strengths (strong $\lambda$ 5780 and $\lambda$ 6283 DIBs) toward SN2001el in NGC~1448 suggest a strong local UV field \citep{Sollerman05DIB}. On the other hand, observations of the ISM toward SN2006X in M100 \citep{Cox08} indicate a dense environment protected from strong UV radiation as it shows a very weak $\lambda$ 6283 DIB and the presence of di-atomic molecules. The $\lambda$ 6283 DIB towards SN2008D in NGC2770 is slightly stronger (EW $\lambda$ 6283/E(B--V) = 1200\,m\AA{}) than the Galactic average (EW $\lambda$ 6283/E(B--V) = 900\,m\AA \citet{Luna08}), while the $\lambda$ 5780 DIB is a little weaker (EW $\lambda$ 5780)/E(B--V) $\sim$\,350\,m\AA{}) than the Galactic average (EW(5780)/E(B-V) = 460~m\AA{}).
For single clouds the $\lambda$ 5780 and $\lambda$ 6283 DIB strengths show a reasonable correlation though variations in their ratios are common. Together with a slightly above average $\lambda$ 5797/$\lambda$ 5780 ratio and the presence of multiple cloud components in the atomic lines this indicates that with SN2008D we probe an ensemble of diffuse clouds with varying local conditions.

For example, SN2006X probes a compact dense interstellar cloud with E(B--V)
$\sim$~1 mag and shows a relative weak $\lambda$ 6283 DIB of 177~m\AA{} (compared to the Galactic average of 900~m\AA{} per E(B--V)) \citep{Cox08, Luna08}. The line-of-sight toward SN2008D instead probes at least two diffuse components and shows a relative strong $\lambda$ 6283 DIB (EW\,=\,850 m\AA{} or EW/E(B--V)\,=\,1200~m\AA) even slightly stronger than the Galactic ``average'' relation EW(5780)/E(B-V) $\sim$~350~m\AA{} for SN2008D, and 460~m\AA{} for MW, and 189~m\AA{} (EW/E(B--V)\,=\,630~m\AA) for SN2001el \citep{Sollerman05DIB}. The $\lambda$ 5780 DIB carrier is sensitive to UV radiation and becomes stronger when exposed to UV radiation. The $\lambda$ 5780 is relatively weak with respect to the Galactic average, while the 6283 is stronger. Although generally these two DIBs show some correlation ratios that clearly differ between different sightlines.

\begin{figure}
\includegraphics[width=\columnwidth]{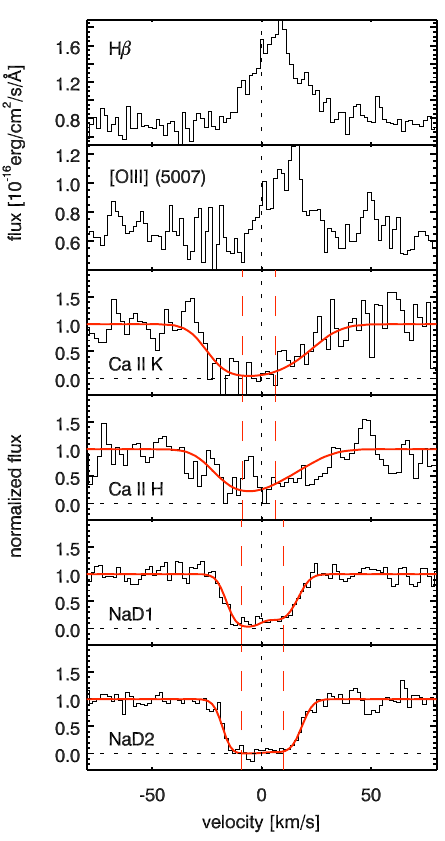}
\caption{Fits to the Calcium and Sodium absorption lines in the UVES spectra and comparison to H$\beta$ and [O\,{\sc iii}] emission lines. The dashed horizontal lines indicate the continuum level. Zero velocity corresponds to the center of the galaxy.
\label{NGC2770:UVESfit}}
\end{figure}

\begin{figure}
\includegraphics[width=7cm]{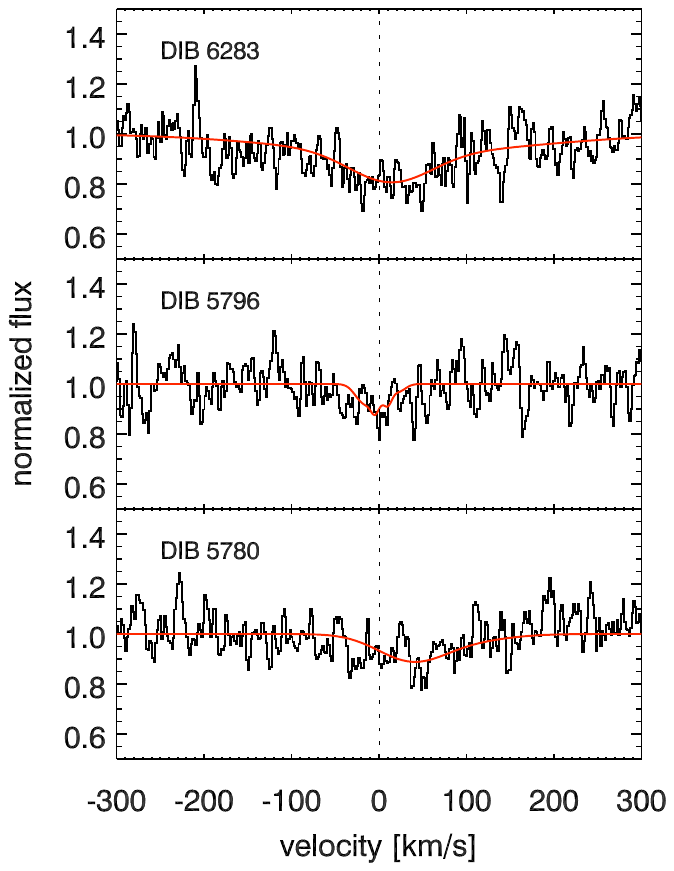}
\caption{Fits to the DIBs detected in the UVES spectra of SN 2008D. Zero velocity corresponds to the center of the galaxy.
\label{NGC2770:DIBs}}
\end{figure}

\subsection{Star formation rates and ages of the stellar population}\label{NGC2770:Halpha}

The local as well as the global SFR in NGC 2770 was determined by using the H$\alpha$ flux and the conversion by \cite{Kennicutt92}. The SFRs at the four different slit positions are given in
Table~\ref{NGC2770:local} as SFR\,kpc$^{-2}$.  We did the same for a range of
regions within NGC 2770 by using the image taken with the H$\alpha$ filter with the continuum subtracted using the offband filter as described in Section~\ref{NGC2770:observations}. 
The SFR at the sites of SNe 2008D and 1999eh are comparable but not especially high.
SN 2007uy lies a few arcsec from two sites with higher SF. The local SFRs elsewhere in NGC 2770 vary from 2$\times$10$^{-3}$ to 4$\times$10$^{-1}$ M$_\odot$~yr$^{-1}$~kpc$^{-2}$. The total SFR
from the H$\alpha$ image taking an elliptic aperture to include the entire galaxy yields a flux of 6.1$\times$10$^{-12}$ erg\,cm$^{-2}$\,s$^{-1}$ or a SFR of 0.42 M$_\odot$~yr$^{-1}$ which is consistent with other measurements from the UV, IR and radio fluxes.

The H$\alpha$ EW is further correlated with the age of the underlying population at least in the first 10$^6$ to 10$^7$yr \citep[see e.g.][]{Zackrisson01}. Table~\ref{NGC2770:local} also lists the H$\alpha$ EW at the different parts within the four slits. We cannot determine the EW at the sites of SN 2008D and SN 2007uy as the continuum is strongly affected by the presence of the SNe and we can only measure the EW of the nearby sites. We find that the stellar populations close to the SN sites are not among the youngest ones.  This is a bit surprising since the progenitors of SNe Ib are expected to be stars with masses $>$ 30--40 M$_\odot$ whose lifetimes are $<$ 6--7 Myr corresponding to a
H$\alpha$ EW of $>$ 100 \AA{} \citep[see e.g.][]{Sollerman05}. There seems to be a strong gradient (see Fig. \ref{NGC2770:slitplots}) in NGC 2770 towards high EWs in the outskirts of the galaxy whereas the center has rather low EWs corresponding to an older population. However, the EWs near all three SN sites are not particularly high which does not suggest particularily young populations.

\subsection{Kinematics}\label{NGC2770:kinematics}
The UVES high resolution data allow us to study the kinematics in the
sightline towards SN 2008D. We use the FITLYMAN program in MIDAS
\citep{Fontana95} to fit different components to the Na I D1 and D2 as well
as the Ca II H \& K lines and to determine column densities 
(see Table~\ref{NGC2770:UVES}). The absorption lines are best fitted with two
components which are at velocities of $-9.21$ and $10.01$~km~s$^{-1}$ 
for the Na I lines and $-$8.60 and 6.54~km~s$^{-1}$ for the Ca~II lines. 
As v$\,=\,$0\,km~s$^{-1}$, we take the redshift of the center of the galaxy
for which we adopt $z=0.007$ as determined from the ALFOSC spectra
probing the center of the galaxy. The two components are very narrow
and have Doppler $b$ parameters of 7.3 and 8.1~km~s$^{-1}$ for Na~I~D and
14.1 and 18.3~km~s$^{-1}$ for Ca~II, the turbulent $b$ parameter is
very small and was fixed to 1~km~s$^{-1}$. 

We compare the position of the absorption with the emission lines
which are likely to originate in the SF regions where the SN reside. 
The emission lines have a width of around 30 km s$^{-1}$ which is
most likely due to the turbulence in the SF region. There is no velocity shift
between the hydrogen and oxygen emission lines suggesting a common
origin for both elements. The redder component of all absorption lines
is coincident with the position of the emission lines while the blue
component is slightly shifted. This indicates that the absorbing
material is most likely in front of the H~II region of the SN. The
position of the $\lambda$ 5797 and $\lambda$ 6283 DIBs are coincident with the other absorption lines whereas the $\lambda$ 5780 DIB is offset by about 40 km/s. A fit with fixing the center of the line to v$=$ 0 km/s gives almost the same $\chi^2$ and a slightly lower line strength of 230 m\AA{}. There is also an intrinsic scatter of the rest wavelength of DIBs from different surveys which vary by 10--20 km/s. Due to the low line strengths and their intrinsically broad profile, it is not possible to distinguish different velocity components in the DIBs.

\section{NGC 2770 in the context of other GRB and SN hosts}

\subsection{Other galaxies with high SN rate}\label{NGC2770:comparison}

A number of galaxies exist that have had more than one SN detected in the optical. 
The record until today is for NGC 6946 with 9 SNe detected,  and where all that
could be classified were SNe II. NGC 6236 (M 83) had 6 SNe and there are 3 galaxies with 5 SNe (NGC 4303 (M61), NGC 4321 (M100) and NGC 2276 (Arp 25). Furthermore, there are 9 galaxies with 4 SNe detected and 26 with 3 SNe of which one is NGC 2770 (complete up to
May 30, 2008). We compiled this information from the CfA database  including only galaxies with a catalogue name. We also include in our sample of frequent SN galaxies 4 galaxies that had 2 stripped-envelope SNe (to
date NGC 2770 is the only one with 3). This sample is clearly biased but may still be useful for the comparison between frequent SN galaxies.

\begin{figure}
\includegraphics[width=\columnwidth]{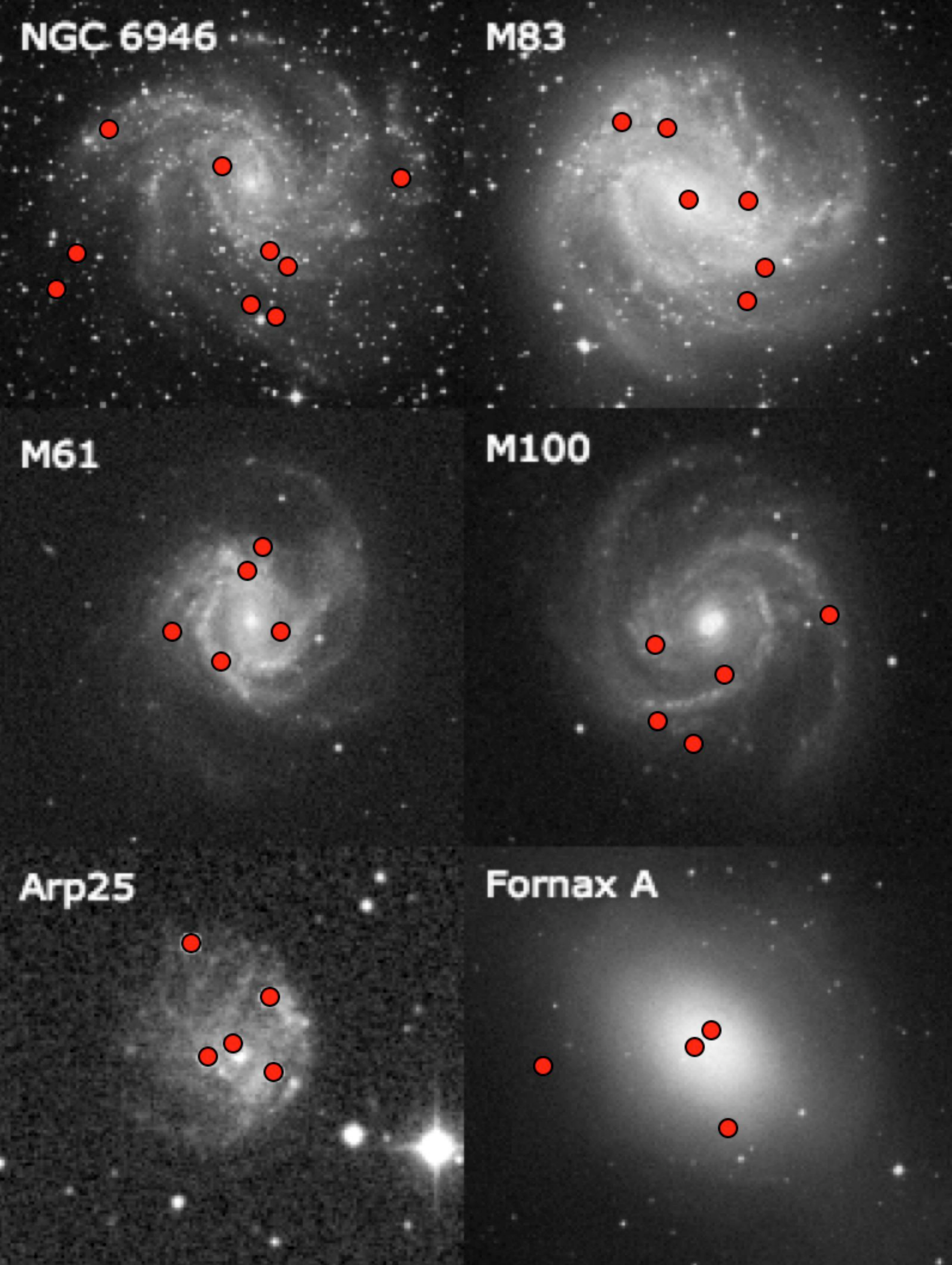}
\caption{Images of the top 5 SN galaxies and Fornax A (only Ia SNe) with the positions of the SNe indicated.
\label{NGC2770:SNgalaxies}}
\end{figure}

In Table~\ref{NGC2770:SNgalaxies} we list the SFRs, SNRs and
specific SFRs weighted by the mass represented by the K-band
luminosity for this sample. The radio SFRs are derived from
\citet[][eq. 15]{YunCarilli02}, the UV SFRs from
\cite{Kennicutt92}. For most galaxies, there is a good agreement
between the UV and radio SFR except for NGC~2276 and 
NGC~3670. Usually, radio SFRs are higher than UV SFRs as the UV is
affected by extinction in the host galaxy, which plays a larger role
in edge on galaxies. In our sample, we do not to account for the extinction
correction and this was not done either for the GRB sample we use for
comparison in Section~\ref{NGC2770:GRBSNhosts}. The SNRs were calculated from
the nonthermal radio flux according to \citet[][eq. 18]{Condon92}
which are based on the assumption that the radio flux comes from
synchrotron radiation from electrons accelerated in the SN remnants.

It may appear surprising that the galaxies with the highest number of
observed SNe do not necessarily have the highest (expected) SNRs as
derived from radio data. However, most galaxies with 4 or more SN detected have an expected SNR between 1--4 in 10 yrs. This is consistent with the number of actually detected SNe considering the small numbers and that SN observations in most of these galaxies have only been possible for
the last few tens of years. NGC 2770 has a SSFR and SNR among the
lowest in our sample and does not reflect the detected SN rate which
could, however, be a statistical fluctuation (see Section \ref{NGC2770:statistics}). 

An important bias in SN detections is the
inclination of the galaxy observed as noted by several authors
\citep[e.g.][]{Cappellaro99}. 
In our sample, the average number of SNe detected per galaxy drops towards higher inclinations, however, the distribution peaks around 30 deg inclination, most likely due to the low number of systems with lower inclinations. 
All of the galaxies with 5 and more SNe detected are seen nearly face-on. If we
would have seen them edge-on, parts of the SNe might have been missed
due to dust extinction or superposition with the bright central
regions. Both issues are currently being addressed
with SN searches in the IR \citep[e.g.][]{Mattila07, Maiolino02, Mannucci03} and continuous
monitoring of galaxies to perform image subtraction or observations in
the radio \citep[e.g.][]{Cram98}

\begin{deluxetable*}{llllllllll} 
\tablewidth{0pt} 
\tablecaption{Galaxies with frequent SN occurrence}
\tablehead{\colhead{galaxy}&\colhead{incl.}&\colhead{SNe}&\colhead{SN types}&\colhead{gal. type}&\colhead{SFR (rad.)}&\colhead{SFR (UV)}&\colhead{SNR}&\colhead{M$_\mathrm{stellar}$}&\colhead{SSFR}\\
\colhead{}&\colhead{$[$deg$]$}&\colhead{}&\colhead{}&\colhead{}&\colhead{$[$M$_\odot$ yr$^{-1}$$]$}&\colhead{$[$M$_\odot$ yr$^{-1}$$]$}&\colhead{$[$yr$^{-1}$$]$}&\colhead{$[$M$_\odot$$]$} &\colhead{$[$Gyr$^{-1}$$]$}}
\startdata
MW&\nodata&4&\nodata&SBc &1&1&0.0100&5 10$^{10}$&0.02\\ \hline
NGC 6946&29&9&II (6)&.SXT6..\tablenotemark{d}&3.125&3.143&0.0744&2.725$\times 10^{10}$&0.115\\
NGC 5236 {\tiny (M83)}&21&6&Ia (1)&.SXS5..\tablenotemark{d}&7.002&0.278&0.1666&6.506$\times 10^{10}$&0.108\\
NGC 4303 {\tiny (M61)}&25&5&II (4)&.SXT4..&11.17& 10.09&0.2650&7.828$\times 10^{10}$&0.143\\
NGC 4321 {\tiny (M100)}&36&5&Ia (4), II (1)&.SXS4..&6.655& 4.569&0.1579&9.942$\times 10^{10}$&0.067\\
NGC 2276 {\tiny (Arp25)}&21&5&II (2)&.SXT5..\tablenotemark{d}&17.55&\nodata&0.4157&2.374$\times 10^{10}$&0.739\\ \hline
NGC 1316 {\tiny (FornaxA)}&56&4&Ia (3)&PLXS0P\tablenotemark{a}&\nodata&1.764 &\nodata&3.146$\times 10^{10}$&15.93\\
NGC 4038&65&4&Ia (1), Ib/c (1)&.SBS9P. \tablenotemark{b,d}&16.00&\nodata&0.3867&\nodata&\\
NGC 3367&13&4&Ia, Ib/c, II&.SBT5..\tablenotemark{a}&11.30& \nodata&0.2671&5.092$\times 10^{10}$&0.222\\
NGC 6754&60&4&II (1), Ia (1)&.SBT3..&2.551&\nodata&0.0580&5.317$\times 10^{11}$&0.048\\
NGC 5468&24&4&II (1), Ia (2)&.SXT6..&\nodata&\nodata&\nodata&\nodata&\nodata\\
NGC 4725&49&4&uncl.&.SXR2P.&0.393& 1.475&0.0093&8.632$\times 10^{11}$&0.005\\
NGC 3184&15&4&uncl.&.SXT6..&0.275& ---&0.0065&7.864$\times 10^{11}$&0.035\\
NGC 2841&65&4&Ia-p (2)&.SAR3*.\tablenotemark{a}&$<$0.343&0.287&$<$0.0082&2.668$\times 10^{11}$&0.013\\
NGC 3690&40&4&Ib (2), II (2)&.IB9P..\tablenotemark{b,e}&67.10& 0.679&1.5873&7.283$\times 10^{11}$&0.921\\ \hline

NGC 521&\nodata&3&II(1)&.SBR4..&0.819&\nodata&0.0193&1.630$\times 10^{11}$&0.006\\
NGC 664&\nodata&3&II(3)& .S..3*.&3.777&\nodata&0.0888&3.938$\times 10^{10}$&0.096\\
NGC 735&\nodata&3&II(1),Ic(1) &.S..3&0.830&\nodata&0.0196&3.044$\times 10^{11}$&0.027\\
NGC 1058&21&3&Ib/c(1), II(1)&.SAT5..&0.019&\nodata&0.0004&1.142$\times 10^{9}$&0.017\\
NGC 1084&56&3&II(2)&.SAS5.. &8.110&\nodata&0.1925&2.324$\times 10^{10}$&0.349\\
NGC 1097&47&3&II(3)&.SBS3..\tablenotemark{f}&7.180&0.172&0.1707&8.872$\times 10^{10}$&0.081\\
NGC 1365&58&3&II(2)&.SBS3.. &15.23&\nodata&0.3619&1.316$\times 10^{11}$&0.116\\
NGC 1448&79&3&Ia(1), II(2)& .SA.6*/&1.394&\nodata&0.0331&2.051$\times 10^{10}$&0.068\\
NGC 1559&56&3&II(2), Ia(1)&.SBS6..&8.049&\nodata&0.1912&1.841$\times 10^{10}$&0.437\\
UGC 1993&\nodata&3&Ia, II& .S..3..&\nodata&\nodata&\nodata&\nodata&\nodata\\
NGC 2207&54&3&Ib(1), Ib/c(1)&.SXT4P.\tablenotemark{b}&29.69&\nodata&0.7035&6.953$\times 10^{10}$&0.427\\
NGC 2770&72&3&Ib(3)&.SAS5*.& 0.586&0.638&0.0139& 9.821$\times 10^{9}$&0.059\\ 
NGC 3147&31&3&Ia, Ib&.SAT4..\tablenotemark{f}&7.393&\nodata&0.1750&1.512$\times 10^{11}$&0.049\\ 
NGC 3627 {\tiny (M66)}&67&3&Ia, II(2)&.SXS3..\tablenotemark{a,f}&2.441&1.111&0.0581&4.086$\times 10^{10}$&0.060\\
NGC 3631&28&3&II (1)&.SAS5.. &1.148&\nodata&0.0276&1.492$\times 10^{10}$&0.077\\
NGC 3938&14&3&Ic,II&.SAS5.. &0.426&0.816&0.0101&8.586$\times 10^{9}$&0.050\\
NGC 3947&31&3&Ia(1), IIn(1)&RSBT3..&2.070&\nodata&0.0486&5.592$\times 10^{10}$&0.037\\
NGC 4157&90&3&II, uncl.(2)&.SXS3\$/ &1.162&\nodata&0.0276&1.184$\times 10^{10}$&0.098\\
NGC 4254 {\tiny (M99)}&\nodata&3&II(1)&.SAS5..&25.09&12.29&0.5944&1.715$\times 10^{10}$&0.146\\
NGC 4374 {\tiny (M84)}&\nodata&3&Ia(2)&.E.1... \tablenotemark{a,c,f}&\nodata&0.064&\nodata&6.344$\times 10^{10}$&1.104\\
NGC 4527&71&3&Ia(1)& .SXS4..\tablenotemark{a}&5.817&\nodata&0.1380&8.834$\times 10^{10}$&0.066\\ 
NGC 4939&\nodata&3&II(1)&.SAS4..\tablenotemark{f}&4.971&\nodata&0.1177&7.156$\times 10^{10}$&0.069\\
NGC 5033&70&3&II(2)&.SAS5..\tablenotemark{f}&2.775&0.393&0.0660&2.201$\times 10^{10}$&0.126\\
NGC 5253&90&3&I(1)& .I..9P*\tablenotemark{d}&0.148&0.079&0.0035&1.445$\times 10^{9}$&0.102\\
NGC 5457 {\tiny (M101)}&7&3&II(1)& .SXT6..&0.443&0.270&0.0105&6.307$\times 10^{9}$&0.070\\
NGC 5668&22&3&Ia, II& .SAS7..&0.603&\nodata&0.0143&9.036$\times 10^{8}$&0.668\\ \hline

NGC 3810&\nodata&2&Ib(1), Ic (1)&.SAT5..&1.249&\nodata&0.0297&1.134$\times 10^{10}$&0.110\\
NGC 7714&42&2&Ib/c(2)&.SBS3*P\tablenotemark{a,d,e}&1.258& 0.952&0.0298&1.704$\times 10^{10}$&0.074\\
NGC 4568&65&2&Ib(1), Ic (1)&(SAbc)\tablenotemark{b}&6.391&1.128 &0.1514&8.751$\times 10^{10}$&0.073\\
NGC 3464&50&2&Ib(1), Ic(1)&.SBT5..&\nodata&\nodata&\nodata&\nodata&\nodata\\
\enddata
\label{NGC2770:SNgalaxiestable}
\tablecomments{Complete sample of all galaxies with $\geq$ 3 SNe and 5 galaxies with 2 stripped envelope SNe, compiled from the CfA SN database. The classifications are adopted from RC3 \tablenotemark{g}, for those galaxies without RC3 classifications, we use the classification from NED. The adopted T value is the number in the classification code for spirals, the only elliptical galaxy, NGC 1316 has T$=$--2. Radio data, UV flux and K-band luminosity are taken from NED where available. SNRs are derived from radio data according to \citet[][eq. 18]{Condon92}, SFRs are from \citet[][eq. 15]{YunCarilli02} for radio and \cite{Kennicutt92} for UV SFRs. }
\tablenotetext{a}{low ionization nuclear emission line region (LINER)}
\tablenotetext{b}{interacting galaxy}
\tablenotetext{c}{low excitation radio galaxy (LERG)}
\tablenotetext{d}{starburst}
\tablenotetext{e}{Wolf-Rayet galaxy}
\tablenotetext{f}{Seyfert galaxy}
\tablenotetext{g}{1st column: R = outer ring; P = pseudo outer ring; C = compact; D = dwarf, 2nd column: E, S, I, P: elliptical, spiral, irregular, peculiar, 3rd column: X = transition between A (no bar) and B (barred galaxy). "+": transition between elliptical and lenticular galaxies (E/S0); 4th column: T = transition between s (no inner ring) and r (strong inner ring), number = ellipticity for an elliptical galaxy: 0 = round, 6, 7 = edge-on; 0 = I0 galaxies; 5th column: lenticulars: -, 0, or + (lenticular stages), spirals: Hubble stage: 0 for 0/a, 1 for a, ... 9 for m, for Magellanic irregulars, always 9; 6th column: Uncertainty (* or ), P (minor peculiarity), or / = sp (spindle; an edgewise galaxy); 7th column: same as 6th, if needed. }
\end{deluxetable*}

\begin{figure}
\includegraphics[width=\columnwidth]{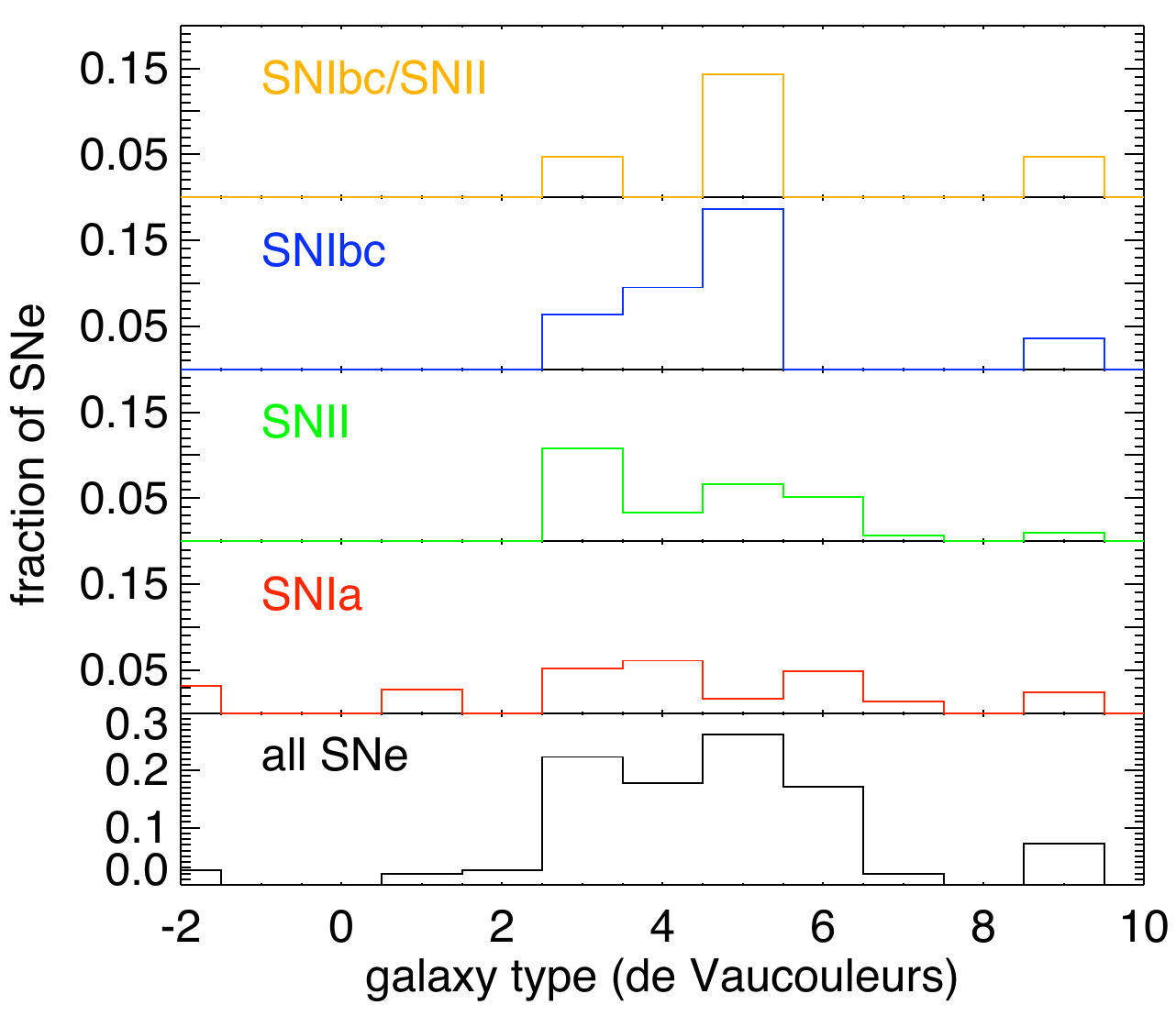}
\caption{Fraction of detected SNe sorted according to the de Vaucouleurs classification \citep{deVaucouleur} of their host galaxies. SNe Ib/c seem to be more concentrated towards later type spirals whereas SNe II seem to prefer earlier type spirals which are also expected to have an older stellar population. In general, most SNe are detected in spiral galaxies. 
\label{NGC2770:SNfraction}}
\end{figure}

It is interesting to note that many of the galaxies with frequent SN
Ib/c occurrence are classified as starbursts, irregular or interacting
with a neighboring galaxy whereas the galaxies with mainly SNe II or
Ia are more ``ordinary'' spirals.  This result is not surprising as
SN Ib/c progenitors are likely more massive than those of SNe II and should
therefore come from a younger population. Interaction between galaxies
is considered as a possible trigger for star formation. 
In Fig.~\ref{NGC2770:SNfraction}, we display the fraction of each SN type which
occurred in the same type of galaxy. For the galaxy classification we
adopted the values from \cite{deVaucouleur} which attributes values
from --\,6 to 11 with --6 to 0 for elliptical galaxies, 1--9 spirals of increasing Hubble type and 10, 11 for irregular galaxies. SNe Ibc seem to be more frequently detected in later type
spirals compared to SN Type II, whereas SN Ia are rather evenly
distributed over all galaxy types. \cite{vandenBergh} investigated the complete sample of SNe detected with the LOSS SN search and their hosts up to 2005 and do not find a significant different between the Hubble Type of SN Type Ib/c and SN Type II. For our sample, we do only consider frequent SN hosts which introduces some bias in the sample. Panel I in Fig. \ref{NGC2770:SNfraction}, however, shows that, at least in our sample, there is a tendency of a higher fraction of SNe Ib/c over SNe II for spirals of later Hubble type.

Furthermore, some galaxies seem to produce only a certain type of
SNe. In order to investigate this notion, we calculated the fraction of
galaxies that only had one SN type detected from our sample of 44
frequent SN galaxies. 29 of those had SNe II out of which 51\%
had only SNe II, 13 had SNe Ib/c with 46\% only having SNe Ib/c
and 17 with SNe Ia of which only 29\% had uniquely SNe Ia. 
In these numbers, we only include the classified SNe. It seems that
indeed, galaxies with SNe II and Ib/c often only have one
type of SN whereas this is less pronounced for SNe Ia.
 This result suggests that the SF history in those galaxies is rather 
uniform in the sense that SF has a common onset in time and first
produces mainly SNe Ibc expected to come from the most massive
stars, whereas later on, when the dominant population has become
older, mainly SNe II are produced. 

\subsection{Comparison to other GRB and SN hosts}\label{NGC2770:GRBSNhosts}

\begin{figure}
\includegraphics[width=\columnwidth]{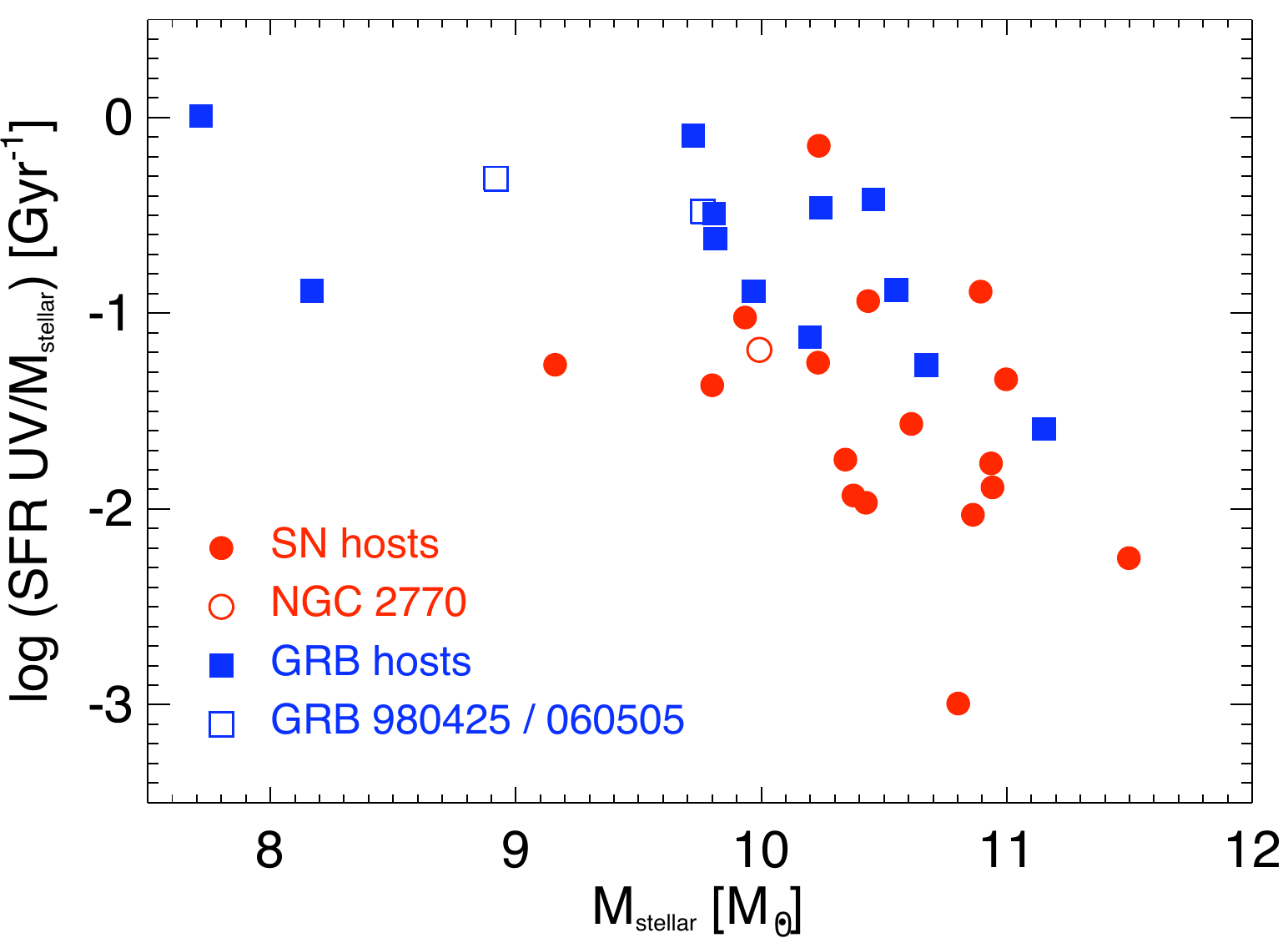}
\caption{Specific SFR from the UV versus stellar mass derived from the
  $K$-band luminosity for GRB hosts (blue squares) and the sample of
  frequent SN galaxies (red dots) presented in
  Tab.\ref{SNgalaxiestable}. The GRB hosts specifically mentioned in
  Sec. \ref{SNandGRBs} and NGC 2770 are plotted as empty symbols.
\label{GRBSNhosts}}
\end{figure}

We now compare our sample of frequent SN galaxies to the hosts of long GRBs 
that are known to be connected to broadlined SNe Ic. For the GRB
hosts, we take the sample from \cite{JMCC08} which lists the UV
SFRs derived from the relations of \cite{Kennicutt92} as well as the stellar masses
derived from the restframe $K$ band luminosity.  GRB hosts have generally lower
masses (log M$_\mathrm{stellar,SN}=$10.71 M$_\odot$, log
M$_\mathrm{stellar,GRB}=$10.36 M$_\odot$) and higher SFRs
(SFR$_\mathrm{GRB}=$2.9 M$_\odot$\,yr$^{-1}$, SFR$_\mathrm{SN}=$ 2.0
M$_\odot$\,yr$^{-1}$) than the frequent SN hosts from our sample. This leads to a higher SSFR for GRB hosts (0.075 and 0.32 Gyr$^{-1}$ for SN and GRB hosts respectively) even though the
two samples overlap partially. We note that the masses derived in \cite{JMCC08} are about two times higher than the masses in the sample \cite{Savaglio08} which partially overlap with the
sample of \cite{JMCC08}. This difference is partly explained by the use of different assumptions for the ratio M$_\mathrm{stellar}$/L$_K$ in both works. NGC 2770 has a SSFR close to the average in the frequent SN host sample, but a somewhat lower mass, but the SSFR lies clearly in the lower end of the
distribution of long GRB hosts.  

The hosts of broadlined  SNe Ic from the sample of \cite{Modjaz08a}
have SFRs comparable to those of GRB hosts, but most likely smaller
SSFRs due to their higher masses. As mentioned above, hosts of broadlined SNe Ic have higher metallicities
than nearby GRB hosts which always have subsolar metallicities, some
of them even down to 1/10 solar \citep[for a complete sample see][and
  references therein]{Savaglio08}. \cite{Modjaz08a} also note that the
properties of the hosts of broadlined SNe Ic are consistent with those
of normal nearby starforming spiral galaxies.

Some nearby GRB hosts are also different from the usual picture of low
mass, high SF GRB hosts.  The host
of GRB 980425, connected to SN 1998bw (broadlined SN Ic) is a spiral
galaxy with not particularly high SF in most parts of the galaxy, 1/3
of the global SF derived from this galaxy comes from a very luminous
Wolf-Rayet (WR) region close to the GRB site \citep{Christensen08}. 
The host of GRB 060505, a long GRB without SN,
is a spiral galaxy and displays relatively high SF and a low
metallicity at the site of the GRB \citep{Thoene08}, but not in the
galaxy in general, even though the SF is still higher than in NGC
2770. It is still an open question whether GRB hosts are a
special population of star-forming galaxies or following the general
trend of other galaxies with redshift.

\subsection{SN detection rates and the probability of finding 3 SNe in NGC 2770 in 10 years}\label{NGC2770:statistics}
NGC 2770 apparently has a high observed SN Ib rate with 3 explosions within ten
years.  Indeed, although 39 more galaxies are known to have had 3 or
more SNe, no other galaxy has had more than two Ib/c SNe. Is this unexpected?
 
To date, no independent measurement of the SN rate of SNe Ib exists. In most cases \citep[e.g.][]{Dahlen04,Cappellaro05} rates in the literature are given for all Core Collapse Supernovae 
(CC SNe) together. Such results are, however, dominated by SNe II,
which are the most frequent SNe. Some papers list at least SNe Ib/c
and SNe II separately \citep[e.g.][]{Cappellaro99,Mannucci05}. 

The rate measurements are either volumetric in case of higher-z SN
surveys \citep[e.g.][]{Dahlen04,SNLSrates}, or galaxy monitoring
type searches, i.e. searches that look repetitively with a given
cadence at a sample of galaxies. This kind of search usually has a
number of biases, such as the frequency of observations, the limiting
magnitudes achieved, the techniques used for detection or the galaxy
inclination, and are restricted to lower or intermediate redshift
\citep{Cappellaro99,STRESS}.

Unlike SNe Ia, where an accurate and complete volumetric SN rate
at low redshift has been measured \citep{Dilday08}, the most recent
results for CC-SNe is still from \cite{Cappellaro99}. Making estimates to
correct for a number of biases, \citep{Cappellaro97} combine
the results of 5 different supernova searches (including pre-CCD data)
and obtain a rate of $0.08 \pm 0.04 \times (\mathrm{H}_\mathrm{o}/75)^2$ SNu for SNe Ib/c
(1 SNu = 1 SN per 100 yr per $10^{10}L_{\odot}^B$). This rate
increases to  $0.14 \pm 0.07 \times (\mathrm{H}_\mathrm{o}/75)^2$ if one only takes Sbc-Sd type
galaxies. For NGC 2770, we find a B-band luminosity from NED of $1.1\times 10^{10}L_{\odot}^B$. Assuming a Poissonian distribution for supernova events, the probability of NGC 2770 to have 3 SNe Ib/c in a decade is then $6.1 \times 10^{-7}$. The probability of getting 3 SN Ib
events is even smaller. 

We cross-check our expected number of SN events in NGC 2770 with
\cite{Mannucci05}, who normalized the SN rate to the K-band luminosity
(SNuK) by using the same sample of SN host galaxies as
\cite{Cappellaro99}. The $K$-band luminosity is a better indicator of the
stellar mass than the $B$-band luminosity which traces mostly the
young star population. With a $K$ band magnitude of 9.57 for NGC 2770 and by comparing with their predicted rates for
Sbc galaxies \citep[][Tab. 2]{Mannucci05} we calculate a probability of
$1.5 \times 10^{-6} $ for NGC 2770 to have 3 SN Ib/c events within a
decade. The two probabilities derived differ by a factor of about two.
However, the chance probability for the observed SN rate at NGC 2770
is overall very small. In addition, the B-K color of NGC 2770 is 3.20,
a value that does not indicate an extreme CC SN production \citep{Mannucci05}. We note that the values derived here are only the probability to detect a SN from statistical considerations and the luminosity of the galaxy and do not necessarily correlate with the expected SN rate from radio observations. 

However, we also have to account for the number of observed galaxies to derive a number for the
probability to find one such galaxy which produced 3 SNe Ib.  
SNe have been discovered in $\sim 3500$ local galaxies but the total number of
\textit{monitored} galaxies is difficult to estimate, due also to the
lack of information and uncertainties in individual amateur
efforts. The results from \cite{Cappellaro99} and \cite{Mannucci05} are based
on samples of 9346 and 8450 galaxies respectively. The LOSS search
\citep{FilipLOSS}, which is the most successful and systematic search
in low-z, monitors around 5000 local galaxies. 
If we now assume $\sim 10000$ monitored galaxies, the chance of observing this
high SN rate in any galaxy becomes $0.6 - 1.5 \%$ which is still a fairly 
low probability but given the uncertainties it 
could be consistent with a chance coincidence. This
would also support the conclusions derived in Chapter~\ref{NGC2770:global} and
Chapter \ref{NGC2770:comparison} that NGC 2770 has no extraordinary global
properties. 

As a curiosity we add here that the same method of calculating
probabilities, can to some extent explain the observed SN rates in the
top SN producing galaxies (those with more than 5 observed SNe) with
probabilities in the range $10^{-3}-10^{-1}$, except the top SN
producing galaxy, NGC 6946, which has a SED very similar to NGC 2770
(chance probability lower than $10^{-4}$).

\section{NGC 2770B}
We also obtained spectra of the neighboring galaxy, NGC 2770B, which
has the same redshift as NGC 2770 \citep{FynboNGC2770GCN} and can therefore
be considered as a companion of NGC 2770. The galaxy consists of two
``blobs'' and we divide the 2D spectrum in 3 parts of which two are
part of the first ``blob'' but show a small spatial separation in the
emission lines in the 2D spectrum. Both regions shows very strong
emission lines which points to a very young population undergoing
heavy star formation. In Table~\ref{NGC2770:NGCB} we list the emission lines
detected as well as their fluxes.

\begin{figure*}
\centering
\includegraphics[width=15cm]{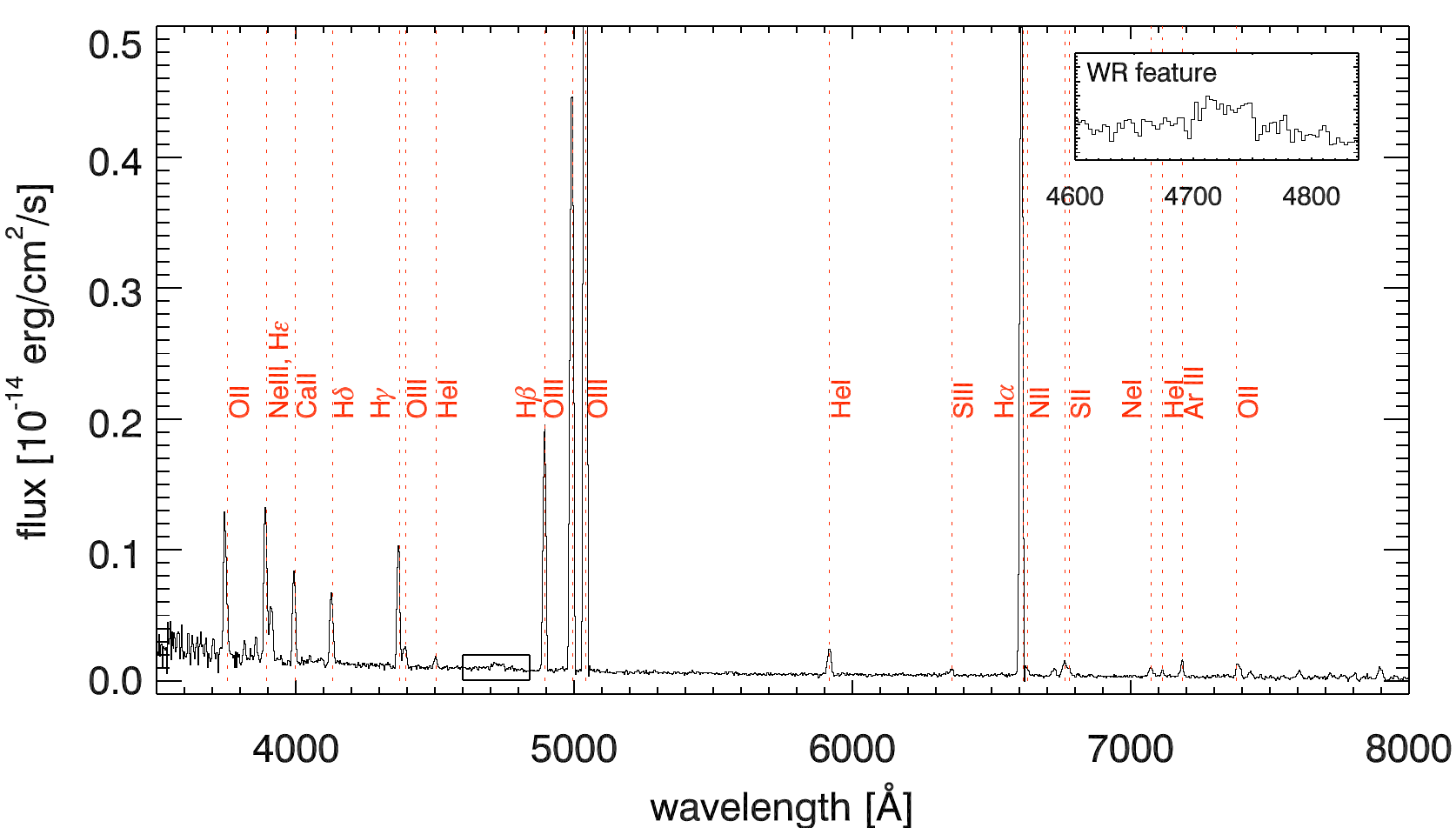}
\caption{Spectrum of the region 1 in NGC 2770B which has the strongest emission lines. The emission lines of [O\,{\sc iii}] and H$\alpha$ have been cutted in order to enlarge the scale to show the weaker emission lines. 
\label{NGC2770:NGC2770Bfig}}
\end{figure*}

We detect a large range of emission lines not detected in the spectra of NGC 2770 including [Ne\,{\sc iii}], H$\delta$, Ca\,{\sc ii}, Na\,{\sc i}, [O\,{\sc iii}] $\lambda$ 4363, He\,{\sc i}, He\,{\sc ii}, [S\,{\sc iii}] and [Ar\,{\sc iii}] where the western ``blob'' of the galaxy shows most emission lines. [Ne\,{\sc iii}] $\lambda$ 3869 is exceptionally strong, especially in the western blob which indicates a large number of young stars. The spectrum resembles very much the nebular spectrum of young HII regions such as the Orion nebula in the MW. In this region, we also detect a clear Wolf-Rayet (WR) feature at 4640 \AA{} (restframe). In Fig. \ref{NGC2770:NGC2770Bfig}, we show the spectrum of region one which has the strongest emission lines with an inset showing an enlargement of the WR feature.

\begin{deluxetable}{llll} 
\tablewidth{0pt} 
\tablecaption{Line fluxes in NGC 2770B}
\tablehead{\colhead{Line} &\colhead{region 1} & \colhead{region 2} &\colhead{region 3}\\
\colhead{} &  \multicolumn{3}{c}{flux $[$10$^{-15}$erg\,cm$^{-2}$\,s$^{-1}]$} }
\startdata
$[$O\,{\sc ii}$]$ $\lambda$ 3729	&13.8	&7.07	&5.95	\\
$[$Ne\,{\sc iii}$]$ $\lambda$ 3869 	&15.5	&2.43	&1.71	\\
H$\epsilon$ 	&6.47		& \nodata	& \nodata	\\
Ca\,{\sc ii} $\lambda$ 3969	&9.69	&0.73	&1.06	\\
H$\delta$			&7.34	&1.16	&1.17	\\
H$\gamma$		&14.3	&1.46	&1.74	\\
$[$O\,{\sc iii}$]$ $\lambda$ 4363	&3.23	&\nodata	&0.28	\\
HeI  $\lambda$ 4472		&1.02	&\nodata	&\nodata	\\
H$\beta$			&29.6	&4.37	&4.04	\\ \hline
EW$[$\AA$]$		&615	&44.4	&119	\\ \hline
$[$O\,{\sc iii}$]$ $\lambda$ 4960	&71.1	&8.88	&7.81	\\
$[$O\,{\sc iii}$]$ $\lambda$ 5008	&194	&25.8	&23.4	\\
He\,{\sc i} $\lambda$ 5876		&3.11	&0.13	&0.49	\\
$[$S\,{\sc iii}$]$ $\lambda$ 6313			&0.84	&0.23	&\nodata	\\
H$\alpha$		&71.3	&14.1	&12.6	\\ \hline
EW$[$\AA$]$		&2200	&295	&709	\\ \hline
$[$N\,{\sc ii}$]$ $\lambda$ 6585	&0.51	&0.04	&0.14	\\
$[$S\,{\sc ii}$]$ $\lambda$ 6718	&0.97	&0.54	&0.30	\\
$[$S\,{\sc ii}$]$ $\lambda$ 6732	&2.60	&0.19	&0.89	\\
Ne\,{\sc ii} $\lambda$ 7024 		&1.14	&0.68	&0.45	\\
He\,{\sc ii} $\lambda$ 7065		&0.63	&\nodata	&\nodata	\\
$[$Ar\,{\sc ii}$]$ $\lambda$ 7135		&1.76	&0.33	&0.29	\\
$[$O\,{\sc ii}$]$ $\lambda$ 7330	&1.81	&\nodata	&\nodata	\\ \hline
12+log(O/H)		&7.19	&7.29	&7.26	\\
E(B--V)			&0		&0	&0	\\
SFR arcsec$^{-2}$	&0.025	&0.0049	&0.0043	
\enddata
\label{NGC2770:NGCB}
\tablecomments{Fluxes and properties for three regions in NGC 2770B. The metallicity has been derived from the R$_{23}$ parameter, the extinction from the Balmer line decrement, the SFR from H$\alpha$ and is listed as SFR/arcsec$^2$.}
\end{deluxetable}

The extinction as derived from the Balmer decrement using both H$\alpha$/H$\beta$ as well as $H\beta$/H$\gamma$ is consistent with zero. In order to determine the metallicity for NGC 2770B we cannot use any of the measurements which require [N\,{\sc ii}] that is only barely detected. This might be both an effect of the low metallicitiy and probably also a high temperature in the HII regions. Instead, we use the R$_{23}$ parameter \citep{KD02} based on the line fluxes of H$\alpha$, H$\beta$, [O\,{\sc iii}] and [O\,{\sc ii}] which are all detected with high S/N in contrast to the spectra of NGC 2770. The metallicities we derive are among the lowest detected for nearby galaxies with 1/30 to 1/25 solar.

In two of the three spectra extracted from NGC 2770B, we also detected the [O\,{\sc iii}] $\lambda$ 4363 line which allows us to determine the electron temperature. From the electron temperature, it is possible to derive directly the abundances of O$^{+}$/H and O$^{++}$/H by using the method described by \citet[][eq. 1, 2, 4 and 5]{Izotov06} and assuming that T$_e$(O\,{\sc ii})=--0.577+T$_e$(O\,{\sc iii})[2.065--0.498T$_e$(O\,{\sc iii})] which \cite{Izotov06} suggests for low metallicities (eq. 14). We then get values of T$_e$(O\,{\sc ii}) = 1.4$\times$10$^{4}$ K for region 1 and T$_e$(O\,{\sc ii}) = 1.2$\times$10$^{4}$ K for region 3 which gives metallicities of 12+log(O/H)=7.1 and 7.6 respectively. The value for region 1 agrees very well with the one derived from R$_{23}$ whereas the value for region 3 from the T$_e$ method is higher. We note, however, that the [O\,{\sc iii}] $\lambda$ 4363 line in region 3 was not detected with high significance.

The SFR in the eastern blob is comparable to the SFR in the nucleus of NGC 2770, however in
the western part it is about an order of magnitude higher than in any region in NGC 2770. The very high EW of the H$\alpha$ line also indicate a dominant stellar population of only a few Myrs. We also
calculate the global SFR from collapsing the entire spectrum of NGC 2770B and obtain a SFR of 0.11 M$_\odot$\,yr$^{-1}$. We obtained images from the 2MASS catalogue in $K_s$ band and determined a magnitude of K$_s$ = 15.3 mag for NGC 2770B. Assuming a mass-to-light ratio of 0.4, the median value adopted by \cite{JMCC08} for GRB host galaxies, and taking eq. 1 from \cite{JMCC08}, we determine a mass of only log M$_*$=7.65 M$_\odot$. Together with the SFR derived from H$\alpha$, this gives a very high specific SFR of 0.39 Gyr$^{-1}$.

NGC 2770B is a rather unsual galaxy compared to its massive neighbour NGC 2770. The fact that no SN has been detected (yet) in this galaxy is likely only due to its very low total mass so is the expected observed SN rate. It is also one of the most metal poor galaxies ever detected \citep[see e.g.][]{Izotov06} which further indicates a very young stellar population. What triggered this high SFR in NGC 2770B remains unclear. It is also interesting that the most metal poor region has a clear detection of a WR feature. WR stars do only occur in a very narrow time window after the onset of a starburst between 2 and 5$\times$10$^{6}$ yr and are therefore a good tracer of the SF history. Models for WR stars predict a strong metallicity dependence on the wind which produces these features \citep[e.g.][]{Vink01}. However, there are a number of low metallicity galaxies with detected WR features which have challenged this model \citep[e.g.][]{Brinchmann08}. Possible explanations are that there are actually more WR stars produced than predicted from the models or that WR stars which are rapidly rotating can form at much lower metallicities. Rapidly rotating WR stars have also been suggested as progenitor for GRBs. Overall, NGC 2770B seems to be a exceptional galaxy that is potentially interesting for future studies.

\section{Conclusions}

We have investigated the properties of the three SN sites in NGC 2770, the host of three SNe Ib, and the other regions in the host along 4 longslit positions. Previous observations in all wavelengths from UV to radio allow us to fit the SED of NGC 2770 and derive a range of global properties from it. We then set NGC 2770 in context to a sample of galaxies with frequent SN occurrence (3 or more SNe detected) and also compute the probability to detect three SNe Ib in a galaxy within only 10 years as it was the case for NGC 2770. From this analyses, we then conclude the following: 
\begin{itemize}
\item NGC 2770 has global properties similar to the MW, even though it has a higher number of SNe observed. Its SFR and SNR are around the average of other nearby spiral galaxies and our sample of frequent SN galaxies. 
\item The only outstanding property of NGC 2770 is its high HI mass which indicates a large reservoir for forming stars. 
\item The metallicities at the SN sites in NGC 2770 are around 0.5 solar which is similar to the values observed for nearby GRB sites, but lower than for broadline SN Ic sites.
\item Almost half of the galaxies with SNe II and Ib/c only have one type of SN which is likely connected to the age of the dominant stellar population. Galaxies producing SNe Ib/cs have a higher deVaucouleurs number than those producing SN II.
\item SN and GRB hosts seem to be somewhat different in terms of SFRs and masses.
\item The probability to detect 3 SNe Ib in a galaxy is 0.6 to 1.5 \% assuming 10,000 monitored galaxies.
\end{itemize}
It therefore seems to be likely that observing 3 SNe Ib in NGC 2770 was only a chance coincidence. NGC 2770 is by no means a special galaxy that would be predestined to produce only stripped-envelope SNe. In fact its properties are not typical for galaxies with (frequent) SN Ib/c occurrence. However, it might also imply that the local properties at the SN sites are more important, at least in some galaxies, to produce a certain type of SN than its global properties.

\acknowledgments
CT and PMV want to thank C\'edric Ledoux for the reduction of the UVES spectra.

The Dark Cosmology Centre is funded by the Danish National Research Foundation. We thank the staff and the NOT and the VLT to perform the observations. In this work we made use of the NASA Extragalactic Database (NED). NED is operated by the Jet Propulsion Laboratory, California Institute of Technology, under contract with the National Aeronautics and Space Administration. JS is a Royal Swedish Academy of Sciences Research Fellow supported by a grant from the Knut and Alice Wallenberg Foundation.

\end{document}